\documentclass[acmsmall]{acmart}
\usepackage{xcolor}
\usepackage{multirow}
\usepackage{booktabs}
\usepackage{soul, color}
\soulregister{\cite}7 
\soulregister{\ref}7 
\usepackage{makecell}
\usepackage{setspace} 
\usepackage{bbding}
\usepackage{threeparttable}
\usepackage{wasysym}
\usepackage{colortbl}

\AtBeginDocument{%
  }

\setcopyright{acmcopyright}
\copyrightyear{2023}
\acmYear{XXXX}
\acmDOI{XXXXXXX.XXXXXXX}

\acmJournal{CSUR}
\acmVolume{37}
\acmNumber{4}
\acmArticle{111}
\acmMonth{8}

\begin{document}

\title{Membership Inference Attacks and Defenses in Federated Learning: A Survey}

\author{Li Bai}
\email{baili.bai@connect.polyu.hk}
\author{Haibo Hu}
\email{haibo.hu@polyu.edu.hk}
\author{Qingqing Ye}
\email{qqing.ye@polyu.edu.hk}
\author{Haoyang Li}
\email{hao-yang9905.li@connect.polyu.hk}
\affiliation{%
	\institution{The Hong Kong Polytechnic University}
	\state{Hong Kong}
	\country{Hong Kong}
}

\author{Leixia Wang}
\email{leixiawang@ruc.edu.cn}
\affiliation{%
	\institution{Renmin University of China}
	\state{Beijing}
	\country{China}
}

\author{Jianliang Xu}
\email{xujl@comp.hkbu.edu.hk}
\affiliation{%
	\institution{Hong Kong Baptist University}
	\city{Hong Kong}
	\country{Hong Kong}}

\renewcommand{\shortauthors}{Bai et al.}

\begin{abstract}
Federated learning is a decentralized machine learning approach where clients train models locally and share model updates to develop a global model. This enables low-resource devices to collaboratively build a high-quality model without requiring direct access to the raw training data.
However, despite only sharing model updates, federated learning still faces several privacy vulnerabilities. One of the key threats is membership inference attacks, which target clients' privacy by determining whether a specific example is part of the training set. These attacks can compromise sensitive information in real-world applications, such as medical diagnoses within a healthcare system.
Although there has been extensive research on membership inference attacks, a comprehensive and up-to-date survey specifically focused on it within federated learning is still absent.
To fill this gap, we categorize and summarize membership inference attacks and their corresponding defense strategies based on their characteristics in this setting. We introduce a unique taxonomy of existing attack research and provide a systematic overview of various countermeasures. For these studies, we thoroughly analyze the strengths and weaknesses of different approaches. Finally, we identify and discuss key future research directions for readers interested in advancing the field. 
\end{abstract}

\ccsdesc[500]{Security and privacy~Privacy protections}

\keywords{Membership inference attacks, federated learning, deep leaning, privacy risk}


\maketitle

\section{Introduction}
\label{sec:intro}
With the increasing availability of extensive datasets, machine learning (ML) has emerged as a critical technology, facilitating significant advancements across various domains, including computer vision~\cite{ZWC+03, LAS+17, DJD+09, HGL+17, HKZ+16}, natural language processing~\cite{KJD+19, SWJ+21, MTC+13, SIV+14}, and more.
Notably, legal regulations such as the General Data Protection Regulation (GDPR)~\cite{GDPR} and the California Consumer Privacy Act (CCPA)~\cite{CCPA} establish key guidelines for data sharing between organizations and mandate the safeguarding of users' privacy when utilizing such data.
Federated learning (FL), as proposed in~\cite{MBM+17}, is a distributed machine learning paradigm that enables multiple clients to collaboratively train a machine learning model (\textit{global model}) without directly sharing their private data. This approach offers a practical solution to overcome these privacy constraints.
Unlike centralized machine learning, which relies on data aggregation, FL allows training samples to remain local across various organizations or mobile devices. This approach not only enhances the volume of available training data but also supports large-scale training processes. Meanwhile, by decoupling model training from direct access to raw data, FL enables each client to maintain their training data locally,  ensuring compliance with current legal regulations and helping safeguard data privacy.

While FL is designed to be privacy-aware by preventing direct access to raw training data, it remains susceptible to significant privacy risks. Researchers have demonstrated that adversaries can exploit model updates in FL to reconstruct the original training data and labels~\cite{ZBM+20, ZLL+19, SA+18}, infer properties of other clients' training data~\cite{SMW+20, MLS+19}, and even generate representative samples~\cite{HBA+17, FMJ+15, WZS+19}.
Among these privacy risks, membership inference attack (MIA) represents a fundamental privacy violation, which seek to determine whether a specific record is part of the training dataset~\cite{HNS+08, SRM+17, PAT+18, YHL+22}.
Compelling applications for MIAs include: 
1) Privacy breach: MIA can reveal sensitive details about the training data of machine learning models to potential attackers. For example, if an adversary determines that a medical record was used to train a cancer prediction model, they may infer that the individual has cancer.
2) Data censorship: MIA serves as an effective tool for auditing data privacy and compliance. For instance, under the legal requirement of the \textit{right to be forgotten}, MIA can be employed to verify whether a platform has successfully erased a specific data point following a deletion request. 
3) Foundation of advanced attacks: MIA can act as a foundational step in strengthening more sophisticated privacy attacks. Attackers can refine their strategies to explore the target model by determining which samples were used as the training data, such as model extraction attacks~\cite{XYT+22}.
A considerable body of research is dedicated to MIAs specifically crafted for the FL environment.
To name a few, the study~\cite{NMS+19} first proposes an inference algorithm to deduce membership information by exploiting gradients, hidden layer output, and sample loss during the learning process.
Since then, the research community has attempted to extend it to various domains, such as classification models \cite{ZJZ+20, PAM+20, HHS+21, GYB+22}, regression models~\cite{GUS+21} and recommender systems~\cite{LZL+22}, by either exploiting exchanged model update \cite{NMS+19, JLN+23, PGR+22} or the trend of model outputs \cite{ZOX+21, GYB+22, SAK+22, HHS+21}.

Although related work explores MIAs and defenses in both centralized and federated settings, there are significant differences between these approaches, as shown in Table~\ref{tab:diff}. These differences motivate us to focus specifically on works within the federated learning framework.
Considering a centralized learning (CL) setting, where the training data is gathered on a single server~\cite{TAZ+22, DGK+20}, machine learning models are trained using the aggregated dataset and then released to the public.
The differences in MIAs and defensive mechanisms between CL and FL settings are as follows.
\begin{itemize}
\item[1]{{\bf Attacker/Defender role}: 
The roles of adversaries and defenders differ between two settings. In CL, most MIAs are usually conducted by model consumers who primarily have access to model outputs~\cite{HHZ+22}. In FL, however, potential attackers mainly stem from insiders, including the central server and other clients.
In the case of defenses against CL-related MIAs, the responsibility for privacy protection lies with the model owners~\cite{HLY+23}. However, both the central server and clients can implement defensive strategies to prevent membership information leakage in FL~\cite{MHB+18, GRC+17}.
}
\item[2]{{\bf Attack/Defense phase}: 
The phase of when an adversary launches an attack or when a defense algorithm is applied varies. In CL, most MIAs take place during the inference phase, after the target model has been released. In contrast, MIAs in FL focus on the training phase, attempting to compromise membership privacy during the entire convergence process.
In terms of defense, both centralized and federated settings aim to protect data privacy during the training stage. However, the key distinction is that centralized defenses also focus on safeguarding the model's output during the inference stage.
}
\item[3]{{\bf Adversary knowledge}:
Attackers in FL can gain more detailed adversarial knowledge compared to those in CL. MIAs in CL can occur in two settings: in the white-box setting, where attackers have access to the target model and intermediate layer computations, or in the black-box setting, where only the model’s outputs, such as prediction scores~\cite{SRM+17, SAZ+18, SAP+21} or labels~\cite{LZZ+21, CCC+21}, are available for analysis.
In contrast, adversaries in FL have extensive access to the target model's gradients, intermediate computations, and final outputs throughout the learning process~\cite{NMS+19}. Furthermore, FL attackers can closely observe the model’s convergence, allowing them to access multiple historical versions of the target model.
}
\item[4]{{\bf Active strategy}:
	The proactive methods used by attackers to infer membership privacy differ across settings. In FL, adversaries with legitimate access to the training process can maliciously manipulate models, enabling them to conduct powerful MIAs through model poisoning~\cite{NMS+19, LKF+20}. In contrast, CL attackers may poison the data to amplify membership information leakage~\cite{TFS+22, CYS+22}.
}
\item[5]{{\bf Protection core}:
	Since the types of private information at risk of being leaked vary, the targets of defense mechanisms across settings also differ. In CL, defenses aim to make the model outputs indistinguishable between member and non-member samples. In contrast, FL countermeasures focus on protecting model updates from privacy violations by the central server or eavesdroppers, while also safeguarding the model from potential breaches by curious clients.
}
\end{itemize}

\begin{table}[htb]
	\setlength\tabcolsep{6.0pt}
	\centering
	\footnotesize
	\caption{A comparison of membership inference attacks and defenses in centralized and federated learning.}
	\label{tab:diff}
	\begin{threeparttable}
		\begin{tabular}{l|l|l|l}
			\toprule
			& Aspect & Centralized Learning & Federated Learning \\
			\hline
			\hline
			\multirow{8}{*}{\makecell{Membership \\ Inference \\ Attack}} & Attacker role & Model consumer & FL server / client / eavesdropper \\
			\cline{2-4}
			& Attack phase & Inference phase &	Training phase \\
			\cline{2-4}
			& \makecell{Adversary knowledge  (data) } & \makecell[l]{Model output \\ Intermediate computation}  & \makecell[l]{ Gradient \\ Model output \\ Intermediate computation} \\
			\cline{2-4}
			& \makecell{Adversary knowledge (model)} &	Target model  &	\makecell[l]{Target model and  historical versions} \\
			\cline{2-4}
			& Active strategy & Data poisoning	 & Model poisoning \\
			
			\hline
			\hline
			
			\multirow{3}{*}{\makecell{Membership \\ Inference \\ Defense}} & Defender role & Data / Model owner & FL server / client  \\
			\cline{2-4}
			& Defense phase & Training / Inference phase &	Training phase \\
			\cline{2-4}
			& Protection core & Model output & \makecell[l]{Model update / Historical models} \\
			
			\bottomrule
		\end{tabular}

	\end{threeparttable}
\end{table}

However, regarding FL-related MIAs, existing surveys provide only incomplete introductions and preliminary discussions~\cite{LQZ+21, KPH+21, YQY+19, LLH+22, LLH+20, ZJL+21, HHZ+22, BNM+21, JMS+20, CHZ+23}, lacking a comprehensive and systematic review.
As an illustration in Table~\ref{tab:related-survey}, earlier research \cite{LQZ+21, YQY+19, ZJL+21, BNM+21, CHZ+23} briefly outline privacy and security issues in the setting of FL, and \cite{LLH+22, LLH+20} emphasize both inference and poisoning attacks within the FL process.
These surveys only mention a few early studies~\cite{MLS+19, NMS+19} and omit more recently published research~\cite{JLN+23, ZGL+24}.
Additionally, the most relevant article~\cite{HHZ+22} provides a comprehensive summary of MIAs in the CL setting, while offering limited works related to FL \cite{MLS+19, NMS+19, LHK+21, ZJZ+20, CJZ+20, HHS+21}.

\begin{table}[h]
	\setlength\tabcolsep{7.0pt}
	\footnotesize
	\centering
	\caption{Existing surveys about membership inference in federated learning.}	\label{tab:related-survey}
	\begin{threeparttable}
	\begin{tabular}{c|c|c|c|c|c|c|c|c}
		\toprule
		\multirow{2}{*}{\bf Survey} & \multirow{2}{*}{\bf Year} & \multirow{2}{*}{\bf Key Topic} & \multicolumn{2}{c|}{\textbf{\makecell{Attacks\tnote{1}}}} & \multicolumn{4}{c}{\textbf{Defenses\tnote{2}}} \\
		\cline{4-9}
		~  & ~ & ~ & \textbf{\makecell{A1}} & \textbf{\makecell{A2}} & \textbf{\makecell{D1}} & \textbf{\makecell{D2}} & \textbf{\makecell{D3}}  &
		\textbf{\makecell{D4}}  \\
		\hline
		\hline
		\cite{LLH+20} & 2020 & \makecell{Privacy and robustness issues of FL} & \CheckmarkBold & \XSolidBrush & \XSolidBrush & \XSolidBrush & \XSolidBrush & \XSolidBrush  \\
		\hline
		\cite{JMS+20} & 2020 & \makecell{Privacy and security attacks in FL} & \CheckmarkBold & \XSolidBrush & \XSolidBrush & \CheckmarkBold & \CheckmarkBold & \XSolidBrush \\
		\hline
		\cite{BNM+21} & 2021 & \makecell{FL vulnerabilities} & \CheckmarkBold & \XSolidBrush & \CheckmarkBold & \CheckmarkBold & \CheckmarkBold & \CheckmarkBold  \\
		\hline
		\cite{ZJL+21} & 2021 & \makecell{Privacy and security threats in FL} & \CheckmarkBold & \XSolidBrush & \XSolidBrush & \CheckmarkBold & \CheckmarkBold & \CheckmarkBold \\
		\hline
		\cite{LQZ+21} & 2021 & \makecell{FL concept and mechanism} & \CheckmarkBold & \XSolidBrush & \XSolidBrush & \CheckmarkBold & \CheckmarkBold & \XSolidBrush \\
		\hline
		\cite{LLH+22} & 2022 & \makecell{Privacy and robustness in FL} & \CheckmarkBold & \XSolidBrush & \XSolidBrush & \CheckmarkBold & \CheckmarkBold & \XSolidBrush \\
		\hline
		\cite{HHZ+22} & 2022 & \makecell{MIAs and defenses of CL} & \CheckmarkBold & \XSolidBrush & \XSolidBrush & \XSolidBrush & \CheckmarkBold & \XSolidBrush \\
		\hline
		\cite{CHZ+23} & 2023 & \makecell{Privacy and  fairness in FL} & \CheckmarkBold & \XSolidBrush & \XSolidBrush & \CheckmarkBold & \CheckmarkBold & \XSolidBrush \\
		\hline
		\cite{HLY+23} & 2023 & \makecell{Defenses against MIAs in CL} & \CheckmarkBold & \XSolidBrush & \XSolidBrush & \XSolidBrush & \CheckmarkBold & \XSolidBrush  \\
		\hline
		This work & - & \makecell{MIAs and defenses in FL} & \CheckmarkBold & \CheckmarkBold & \CheckmarkBold & \CheckmarkBold & \CheckmarkBold & \CheckmarkBold \\
		\bottomrule
	\end{tabular}

	\begin{tablenotes}
		\item[1] A1: Update-based attack A2: Trend-based attack
		\item[2] D1: Partial sharing D2: Secure aggregation D3: Noise perturbation D4: Anomaly detection
		
		\end{tablenotes}
	\end{threeparttable}
	
\end{table}

\begin{figure}[htbp]
	\centering
	\includegraphics[width=4.5in]{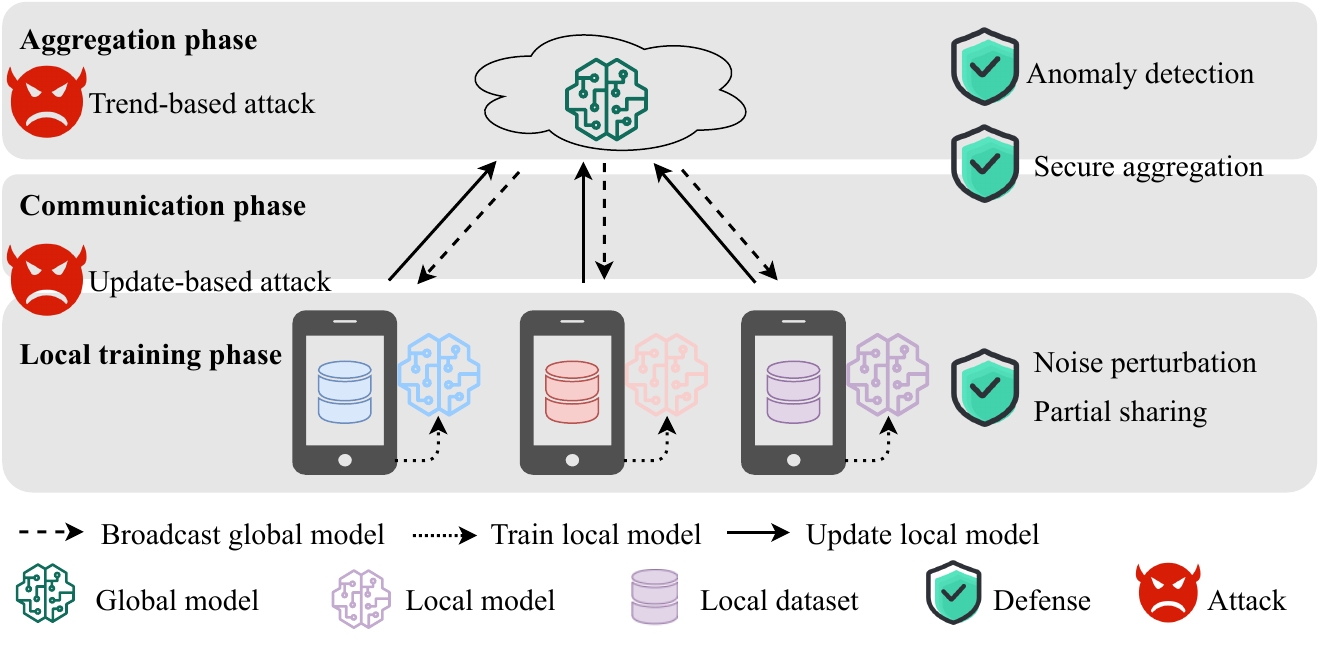}
	\caption{Membership inference attacks and defenses in federated learning.}
	\label{fig:outline}
\end{figure}

In this work, we provide a comprehensive survey of MIAs together with defense strategies on the whole FL process, as shown in Fig.~\ref{fig:outline}.
Starting with a unique taxonomy, we extensively review existing MIAs on FL through model updates and convergence trends in the whole FL training process.
As for defenses against MIAs in the context of FL, we review four mitigation strategies used to protect exchanged updates and models, including partial sharing \cite{SRZ+15, AAH+17}, secure aggregation~\cite{YAC+82}, noise perturbation \cite{ YXF+22, NNJ+22} and anomaly detection~\cite{MMZ+23}.
In the end, we also point out future research directions about MIAs and defenses in FL.
The contributions of this research can be summarized as follows:
\begin{itemize}
\item
We present a comprehensive review of MIAs in the FL setting by summarizing most studies in the literature. To the best of our knowledge, this paper is the first survey of MIAs and defenses in the FL domain.

\item
We identify MIA approaches in the context of FL, and offer both update-based and trend-based perspectives to guide this review. Based on our analysis, we present a unique taxonomy to summarize existing works on MIAs and discuss the differences from CL-related MIAs. 

\item 
We categorize existing countermeasures in FL on four approaches and analyze their advantages and limitations. Additionally, we offer a comparison of defenses in the CL setting from key viewpoints.

\item
We envision promising research directions for MIAs and defenses in this field.
\end{itemize}

The remaining sections of this survey are organized as follows: Section~\ref{sec:preli} first provides the preliminary background and briefly reviews CL-related MIAs and mitigation strategies. Following this, Section~\ref{sec:threat} explores various threat models in FL. Next, Section~\ref{sec:attack} introduces and summarizes existing research on MIAs within the FL context. In Section~\ref{sec:defense}, we describe the countermeasures available to defend against these attacks. Section~\ref{sec:future} then highlights potential research directions, while Section~\ref{sec:concl}concludes with final remarks.

\section{Preliminaries}
\label{sec:preli}

This section first presents background knowledge regarding CL and FL.
Then we offer a brief overview of MIA and defense studies in CL.

\subsection{Centralized Learning}

We consider a centralized setting in which training data are pooled in the server to train an ML model~\cite{TAZ+22, DGK+20}, which leverages input-output pairs to train a non-linear function that predicts outcomes for new queries.
Let $D_{tr}=\left\{\left(\textbf{x}_i,y_i\right)\right\}_{i=0}^{N}$ and $D_{te}=\left\{\left(\textbf{x}_i,y_i\right)\right\}_{i=0}^{M}$ denote the training and test datasets, respectively, where $\textbf{x}_i \in \mathbb{R}^d $ denotes the $d$-dimension feature vector of the $i$-th sample labeled by $y_i$.
The learning goal is to develop an ML model $F$ that maps an input $\textbf{x}$ into $F\left(\textbf{x};\theta\right)$, where $\theta$ is the model parameters.
To obtain the parameters $\theta$, a loss function $\ell\left(\cdot,\cdot\right)$ is typically introduced to measure the discrepancy between the predicted output $F\left(\textbf{x};\theta\right)$ and the ground truth label $y$. 

To obtain a high-quality ML model, we consider the expected value of the loss on the training dataset.
A common approach for model optimization is empirical risk minimization~\cite{VV+91}, which minimizes the following objective function on $D_{tr}$:
\begin{equation}
	\small
	\label{eq:emp}
		L\left(D_{tr}, \theta\right) = \frac{1}{N}\sum_{i=0}^{N} \ell \left(F\left(\textbf{x}_i;\theta\right), y_i\right)
\end{equation}
Many training algorithms have been proposed to minimize this objective function \cite{SD+98, DJH+11, LDP+14}, including stochastic gradient descent (SGD)~\cite{SD+98}, which updates the ML model in the $t$-th iteration as follows:
\begin{equation}
	\small
	\label{eq:sgd}
	\theta^{t} = \theta^{t-1} - \eta \sum_{\left(\textbf{x}_i,y_i\right) \in B}{\nabla_{\theta}\ell \left(F\left(\textbf{x}_i;\theta^{t-1}\right), y_i\right)},
\end{equation}
where $B$ is a mini-batch of random training examples from $D_{tr}$, $\theta^{t}$ represents the model parameters in the $t$-th round, and $\eta$ denotes the learning rate.
The optimization process is terminated when the model converges to a local minimum, or the iteration reaches a preset number. The trained ML model often uses the accuracy on test dataset $D_{te}$ to validate its performance.

\subsection{Federated Learning}
\subsubsection{Brief Introduction of Federated Learning}
FL is a collaborative ML paradigm that is widely used in various applications, such as smart healthcare \cite{XJG+21, NDC+22} and the Internet of Things \cite{KLU+21, NDC+21}.
In contrast to CL, in which training data are collected and processed in a centralized location,
FL involves training a shared global model using distributed data from different organizations or devices.
Instead of transmitting the raw data to a central server, FL allows local data to remain stored locally and only the model parameters or gradients are exchanged, which addresses the data island problem and alleviates data privacy concerns.

The FL system involves two entities, clients and the central server.
The clients (a.k.a., participants) possess training data and collaborate to train a shared model.
According to the number of clients involved, there are two main FL types: cross-silo and cross-device FL~\cite{KPH+21}. Cross-silo FL involves a relatively small number of clients, typically organizations or data centers with a significant amount of data \cite{HYC+21, MOX+20}, while cross-device setting involves a large number of clients with small amounts of data, such as mobile devices \cite{KLU+21, NDC+21}.
The central server is used to aggregate model updates from clients without accessing their local data. In the cross-silo setting, a client can be selected as the central server to perform aggregation and update the global model. In contrast, a powerful server is often introduced to manage model aggregation effectively in cross-device FL.

FL typically involves training a joint global model based on distributed local data through three phases: training, communication, and aggregation phase.
There are $K$ clients $\left\{c_1,c_2,...,c_K\right\}$ that involve in FL and each client $c$ owns a local dataset $D_c$. We use $\theta_{s}$ and $\theta_{c}$ to denote the global and local model, respectively.
To train a global model collaboratively, the central server first initializes the global model $\theta_{s}$ (e.g., model weights and hyperparameters) and broadcasts it to clients, and then the FL process carries out as follows:
\begin{itemize}
	\item[1]{
		Local training phase.
		Each selected client $c$ receives the global model $\theta_{s}$ and training settings. Then, the client fine-tunes it on the local dataset $D_c$.
	}
	\item[2]{
		Communication phase.
		For each selected client, it uploads the latest version of local model $\theta_{c}$ to the central server,
		while the central server broadcasts the global model to selected clients after the aggregation phase.
	}
	\item[3]{
		Aggregation phase.
		The central server aggregates all the model updates from the selected clients and renews the global model $\theta_{s}$.
	}
\end{itemize}
Three phases are repeated until the loss value of the global model converges on a validation dataset, resulting in a well-trained model.

\subsubsection{Categorization of Federated Learning}
Based on the distribution of data over the sample and feature spaces, we can broadly categorize FL into horizontal FL, vertical FL, and federated transfer learning \cite{LQZ+21, KPH+21, YQY+19}.

\begin{figure}[htbp]
	\centering
	\includegraphics[width=4.0in]{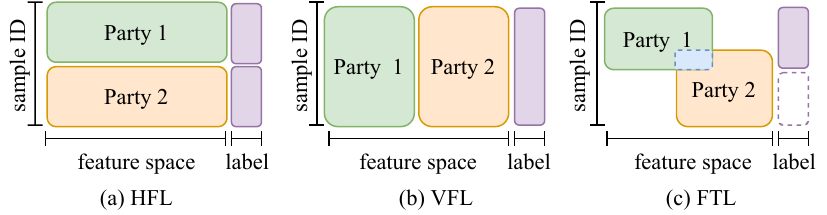}
	\caption{Three categories of federated learning.}
	\label{fig:hfl}
\end{figure}

\textbf{Horizontal Federated Learning (HFL)} refers to scenarios where multiple parties possess data samples with the same features but unique sample IDs, as shown in Fig.\ref{fig:hfl}(a). For instance, different hospitals may possess patient records that share the same feature space but have different patient IDs. HFL is the most popular FL category in the literature \cite{MBM+17, LQZ+20}, and local models commonly share the same architecture as the global model. As such, the global model can be simply aggregated from local models.

\textbf{Vertical Federated Learning (VFL)} refers to scenarios where multiple parties in an FL system hold data samples with identical sample IDs but different feature spaces, as shown in Fig.\ref{fig:hfl}(b).
A typical case for VFL is when an E-commerce company and a local bank collaborate to train a personalized loan model based on online shopping records and credit situation~\cite{YQY+19}. 
Typically, only one participant can access the labels, and sample alignment is performed before training a joint model across multiple clients~\cite{ZJL+21}.

\textbf{Federated Transfer Learning (FTL)} pertains to the situation where clients' datasets contain varying ID spaces and feature spaces~\cite{LYK+20},  as shown in Fig.\ref{fig:hfl}(c). Inspired by transfer learning, FTL allows knowledge to be shared without compromising data privacy, enabling the transferability of complementary knowledge by exploiting source domain knowledge to train an FL model for the target domain \cite{GDL+19, SSX+19}.

\subsection{Membership Inference Attacks in  Centralized Learning}
\label{sec:mia}

MIAs aim to infer whether an example was used to train an ML model~\cite{SRM+17}.
Given a query example $\textbf{x}$ and a target model $F(\theta)$, an MIA algorithm $\mathcal{A}$ infers the membership status $m$ of $\textbf{x}$.
We formulate an MIA algorithm as a binary classification task:
\begin{equation}
	\small
	\label{eq:mia}
	m=\mathcal{A}\left(\textbf{x}, F(\theta)\right) = 
		\begin{cases}
				& 0, \quad \textbf{x} \notin D_{tr} \\
				& 1, \quad \textbf{x} \in D_{tr},
			\end{cases}
\end{equation}
where the membership status is $1$ if the adversary infers that the input $\textbf{x}$ was used to train the target model and $0$ otherwise.
MIAs in CL occur in either \textit{white-box} scenarios, where attackers access model details (e.g., architecture and parameters), or \textit{black-box} scenarios, where only model outputs are observed. In white-box settings, model parameters $\theta$ are observable, but not in black-box settings~\cite{HHS+21}.
Based on the construction of the attack model, MIAs can be categorized into classifier-based attacks and metric-based attacks.
\begin{figure}[htbp]
	\centering
	\includegraphics[width=4.5in]{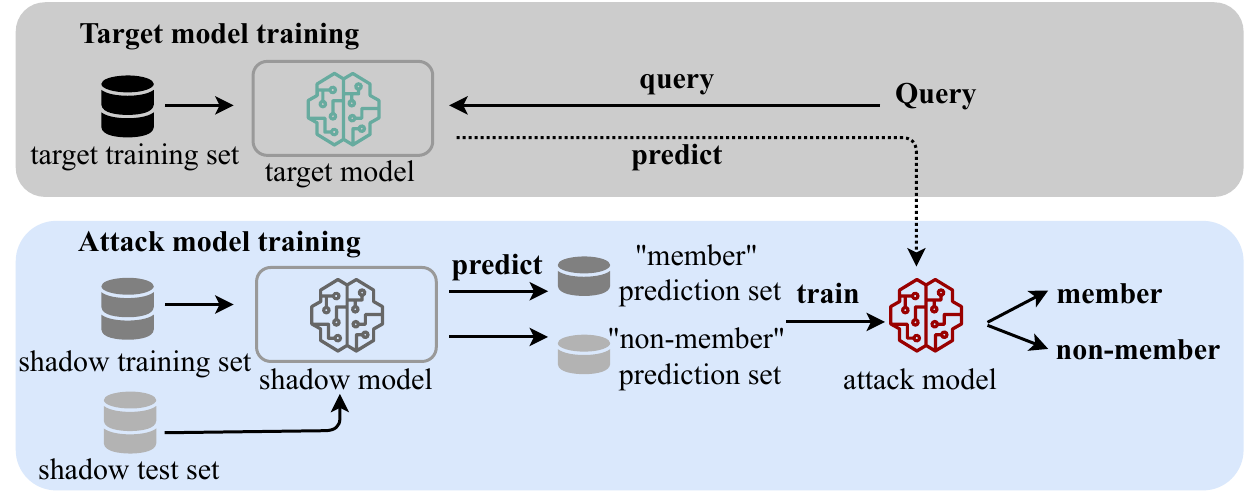}
	\caption{Overview of the shadow training scheme in CL.}
	\label{fig:shadow-train}
\end{figure}

In a classifier-based MIA, a binary classifier attack model is constructed to determine membership information.
Shokri et al.~\cite{SRM+17} conduct the pioneering study on this approach, utilizing the \textit{shadow training} technique to build an attack model that outperforms random guessing (i.e., 0.5 probability of inferring membership).
Fig.\ref{fig:shadow-train} visualizes how shadow training can be used to construct a classifier-based MIA. Assuming the attacker has auxiliary knowledge, they can collect a shadow dataset whose distribution is similar to the target dataset. Next, the attacker trains one or multiple shadow models with the same architecture as the target model and collects prediction vectors from the shadow dataset. Finally, the attacker uses these prediction data to train an attack model capable of inferring membership status based on the output of a query example.
Later studies extended this technique to relaxed assumptions~\cite{SAZ+18} and various model architectures \cite{TSB+19, LKS+19, GUS+21, LHJ+21}.

A metric-based MIA is more straightforward and has less computational cost than classifier-based attacks.
It deduces membership information for data records by calculating metrics on their prediction vectors. The calculated metrics are then compared with a preset threshold to determine the membership status of a data record~\cite{HHZ+22}.
Several metric options are available, such as prediction correctness \cite{YSG+18}, prediction loss \cite{YSG+18}, confidence score \cite{SAZ+18}, and confidence entropy \cite{SAZ+18, SLM+21}.
Based on these intuitive approaches, advanced approaches have been proposed to calibrate calculated metrics or estimate the prediction distribution by introducing more shadow models \cite{WLG+22, CNC+22} to decrease the false positive rate of MIAs.

\subsection{Membership Inference Defenses in Centralized Learning}
\label{sec:def-cl}

Numerous studies have explored membership inference defense techniques in the context of CL, primarily including regularization, knowledge distillation, differential privacy, and output perturbation, designed to thwart membership privacy violations.
In this subsection, we touch upon these strategies. For a more comprehensive understanding, readers are encouraged to delve into the detailed introduction provided in~\cite{HLY+23}.

Regularization and knowledge distillation effectively defend against MIAs by diminishing the overfitting level of ML models. Existing empirical and theoretical research has underscored the link between the leakage of membership information and the extent of model overfitting~\cite{SRM+17, SAM+19, YSG+18}.
Motivated by this observation, regularization techniques strive to reduce the generalization gap, thus diminishing the vulnerability to MIAs. This category encompasses a range of approaches, including L2-norm regularization, dropout techniques~\cite{SAZ+18}, and model stacking methods~\cite{SA+18}.
Likewise, knowledge distillation techniques~\cite{ZJC+21} encourage a smaller student model to learn from the outputs of a teacher model, rather than relying solely on the original training labels. This approach enhances overall model generalization while simultaneously thwarting attackers from inferring sensitive member data.

Differential privacy serves as a potent defense mechanism in the CL setting. This technique operates by introducing carefully calibrated noise into model gradients, thereby guaranteeing that the presence or absence of individual data points remains indiscernible. While this method inherently guards against MIAs as its theoretical definition, it invariably leads to substantial utility loss when implemented~\cite{SRM+17, JBE+19}. 

In contrast to prior defense techniques that alter input data or the training process, output perturbation represents a post-processing method employed to align the model's output across training and test samples. Typically integrated during the inference phase, this approach is efficient against black-box MIAs in CL. It entails the partial disclosure of confidence scores, including the top-$K$ confidence score~\cite{SRM+17}, the associated predicted label~\cite{LZZ+21, CCC+21}, and noisy confidence scores~\cite{JJS+19}.

\section{Threat Model and Attack Taxonomy}
\label{sec:threat}

\subsection{Threat Model}
\begin{figure}[htbp]
	\centering
	\includegraphics[width=4.0in]{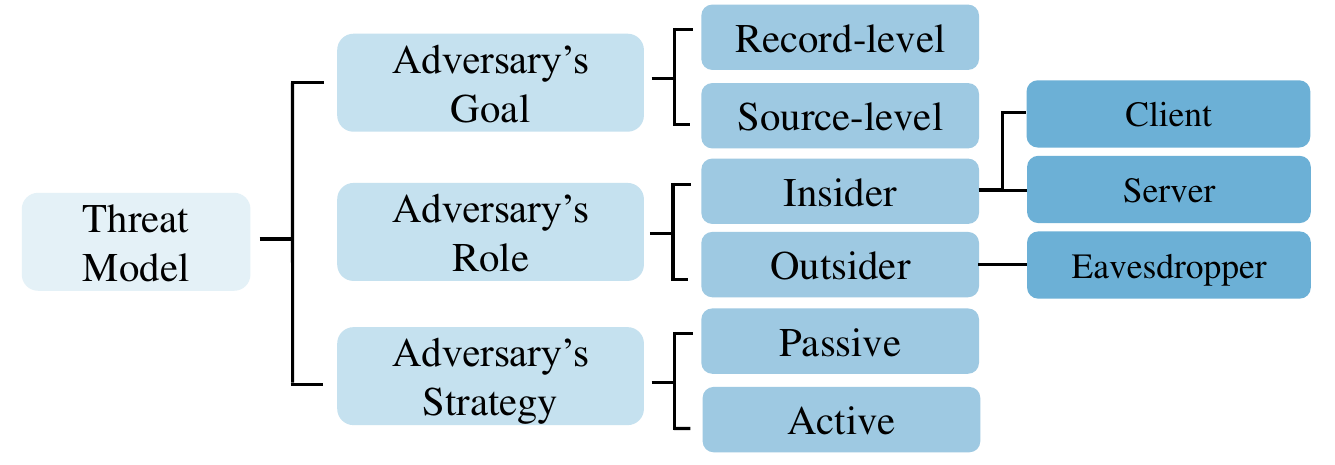}
	\caption{Categorization of threat models.}
	\label{fig:threat}
\end{figure}

In this subsection, we overview the threat models of MIAs in the FL setting from three perspectives: the adversary's goal, role, and strategy, as shown in Fig.\ref{fig:threat}.
In addition, we discuss and compare the adversarial knowledge that an MIA attacker can access in each threat model.

\subsection{Adversary's Goal.}
In the existing literature on MIAs in FL, the granularity of the target can be either \textit{record-level} or \textit{source-level}. As defined in Eq.(\ref{eq:mia}), record-level attacks represent a privacy breach of individual data items.
For instance, in a disease-prediction model trained through FL, record-level MIAs infer a patient who likely suffers from disease by identifying the presence of the patient's clinical record in the \textit{entire training dataset}.

On the other hand, source-level attacks refine the goal of MIAs from the entire training dataset $D_{tr}$ to a \textit{specific client's training dataset} $D_{c}$. 
When a data point is a member in the context of FL, source-level MIAs can further identify the specific client to which it belongs.
Formally, the training dataset consists of multiple local datasets $D_{tr}=\left\{D_c | c=1,2,..., K\right\}$, where $D_c$ is the local dataset of client $c$.
Source-level MIAs aim to trace the source of a training member \cite{HHS+21, CJZ+20, ZYC+21}, formally defined as follows:
\begin{equation}
	\small
	\label{eq:mia2}
	m=\mathcal{A}\left(\textbf{x}, F(\theta)\right) =
	\begin{cases}
		& 0, \quad \textbf{x} \notin D_{tr} \\
		& c, \quad \textbf{x} \in D_c
	\end{cases}
\end{equation}

Source-level attacks are a natural extension of record-level attacks and cause greater privacy concerns~\cite{HHS+21}.
For example, suppose several hospitals collaborate to train a shared global model to predict COVID-19 diagnosis. Record-level MIAs can reveal a patient's identity when her record is used as the training data. By contrast, source-level attacks can identify the hospitals where these patients were treated, which can further leak more information, e.g., patients' addresses.

\subsection{Adversary's Role.} 
ML models in the FL setting are vulnerable to MIAs launched by either \textit{insiders} or \textit{outsiders}.
Insiders, such as curious clients or the central server, may perform MIAs to uncover sensitive information about other participants. Outsiders, like eavesdroppers, can exploit intercepted model updates to infer membership privacy related to either local or global models in the FL system.

As insiders, the FL server and clients can legally access local and global models during training. In particular, a server is more powerful than the clients because it can observe each local model from participants and manipulate the aggregated results. Conversely, clients can only observe the aggregated model, making it challenging to perform source-level attacks without auxiliary knowledge.
While as outsiders, eavesdropping attackers can intercept the communication messages between the FL server and a client, including the global model updates of the server to a client, the local model updates of a client to the server, or both.

In general, insider adversaries pose a more significant threat than outsider adversaries as they can access more information and manipulate the training process~\cite{LLH+22}. For instance, an honest-but-curious FL server may access the local models of each participant during the aggregation process and use this information to identify whether a query sample belongs to a particular participant. A malicious FL participant may even upload a poisoned model to the FL server to enhance sensitive information to leak~\cite{NMS+19}. By comparison, eavesdropping attackers can only passively access communication messages but cannot modify them.

\subsection{Adversary's Strategy.}
An adversary can infer the membership information in the context of FL through \textit{passive} or \textit{active} strategies.

Regarding a passive MIA, an adversary only observes and exploits the learning model without directly interfering, relying solely on the data collected during the normal operation of the FL system.
For example, any client in an FL system can exploit the information they collect during training to deduce private data about other participants.
Such an attack is challenging to detect because the adversary does not affect the target model during the training phase.
In contrast, an insider attacker can launch an active attack to boost attack performance by altering the learning model or data. 
For instance, a malicious client can actively push a target model far away from the optimum to inspect the response of a training member~\cite{NMS+19}. 
Active inference attacks are easier to mount in FL than CL, as any participant or the server can modify the training data or manipulate the model parameters.

\subsection{Attack Taxonomy}

In this survey, we classify existing research on MIAs in FL into two categories: (1) update-based attacks, which leverage one or more historical versions of the target model to infer membership information; and (2) trend-based attacks, which analyze the trajectory of specific indicators to determine membership status.  
Given their different threat models, Table~\ref{tab:newattack} provides a summary of the surveyed attacks.

\begin{table*}[htbp]
	\setlength\tabcolsep{3.0pt}
	\footnotesize
	\centering
	\caption{Taxonomy of membership inference attacks on federated learning.}
	\label{tab:newattack}
	\begin{tabular}{lll|ll|ll|ll}
		\toprule

		\textbf{\makecell{Attack}} & \textbf{\makecell{Approach}} & \textbf{\makecell{Reference}}  & \multicolumn{2}{c}{ \textbf{\makecell{Adversary's \\ Goal}}} & \multicolumn{2}{|c}{ \textbf{\makecell{Adversary's \\ Role}}}  & \multicolumn{2}{|c}{\textbf{\makecell{Adversary's \\ Strategy}}} \\
		\hline
		~ & ~ & ~  & \rotatebox{70}{Record-level} & \rotatebox{70}{Source-level} & \rotatebox{70}{Insider} & \rotatebox{70}{Outsider} & \rotatebox{70}{Passive} & \rotatebox{70}{Active} \\
		\hline
		 \multicolumn{3}{c|}{Update-based MIAs} & & &  &  & &  \\
		\hline
		Model gradient-based  &  Original gradient & Nasr et al.~\cite{NMS+19}  & \checkmark &  & \checkmark &  & \checkmark & \checkmark \\
		~ &  ~ & Gupta et al.~\cite{GUS+21} & \checkmark &  &\checkmark  & & \checkmark &  \\
		~ &  ~ & Lu et al.~\cite{LHL+20} & \checkmark &  & \checkmark &  & \checkmark  &  \\
		\cline{2-9}
		\rowcolor{gray!10}  ~ & Gradient difference & Melis et al.~\cite{MLS+19} & \checkmark  &  & \checkmark &  & \checkmark  &  \\
		\rowcolor{gray!10}  ~ & ~ & Li et al.~\cite{JLN+23} & \checkmark &  & &\checkmark & \checkmark  &  \\
		\rowcolor{gray!10}  ~ & ~ & Zhu et al.~\cite{ZGL+24} & \checkmark &  & \checkmark & & \checkmark & \\
		\hline 
		Single model-based & Shadow training & Liu et al.~\cite{LZL+22}	& \checkmark &  & \checkmark & & \checkmark &  \\	
		~ & ~  & Pustozerova et al.~\cite{PAM+20} & \checkmark &  & \checkmark & & \checkmark  &  \\	
		~ & ~ & Zhang et al.~\cite{ZJZ+20}	&  & \checkmark & \checkmark & & \checkmark  &  \\	
		~ & ~ & Chen et al.~\cite{CJZ+20} &  & \checkmark  & \checkmark & & \checkmark &  \\	
		~ & ~ & Zhao et al.~\cite{ZYC+21}	& \checkmark &  & \checkmark & & \checkmark  &  \\	
		~ & ~ & Luqman et al.~\cite{LAC+23} 	& \checkmark &  &\checkmark & &  & \checkmark \\	
		~ & ~ & Banerjee et al.~\cite{BSR+24}	& \checkmark &  &  & \checkmark & \checkmark &  \\	
		~ & ~ & Truex et al.~\cite{TSL+18}	& \checkmark &  & \checkmark &  & \checkmark &  \\	
		~ & ~ & Yuan et al.~\cite{YWY+23} 	& \checkmark &  & \checkmark &  & \checkmark &  \\			 
		\rowcolor{gray!10} ~ & Structure modifying & Pichler et al.~\cite{PGR+22} & \checkmark &  & \checkmark & \checkmark &  & \checkmark \\
		\rowcolor{gray!10} ~ & ~ & Nguyen et al.~\cite{JHP+22} & \checkmark  &  & \checkmark & &  & \checkmark  \\	
		\hline
		\multicolumn{3}{c|}{Trend-based MIAs} & & &  &  & &  \\
		\hline
		Model output-based &  Prediction trajectory & Zari et al.~\cite{ZOX+21}  & \checkmark &  & & \checkmark & \checkmark &  \\
		~ & ~ & Gu et al.~\cite{GYB+22} & \checkmark  &  & \checkmark & & \checkmark &  \checkmark \\
		~ & ~ & Zhang et al.~\cite{ZYB+23} & \checkmark &  & \checkmark & & &  \checkmark \\
		~ & ~ & Liu et al.~\cite{LGT+23} & \checkmark &  & \checkmark & &\checkmark  &   \\
		\cline{2-9}
		\rowcolor{gray!10}  ~ & Loss trajectory & Hu et al.~\cite{HHS+21} &  & \checkmark  &\checkmark  & & \checkmark &   \\
		\rowcolor{gray!10}  ~ & ~ & Suri et al.~\cite{SAK+22} & \checkmark & \checkmark & \checkmark &  & \checkmark &   \\
		\rowcolor{gray!10}  ~ & ~ & Zhu et al.~\cite{ZGL+24} & \checkmark &  & \checkmark & & \checkmark & \\
		\hline 
		Model parameter-based & Bias trajectory & Zhang et al.~\cite{ZLL+23} & \checkmark &  & \checkmark & & \checkmark  &  \checkmark	 \\	 
		\bottomrule
\end{tabular} 
\end{table*}

\section{Membership Inference Attacks in Federated Learning}
\label{sec:attack}

In this section, we provide a detailed examination of specific attacks within the FL context from both update-based and trend-based perspectives.

\subsection{Update-based Membership Inference Attacks}

Although FL prevents access to the raw training data, it is still vulnerable to MIAs owning to the gradient/weight exchanged.
An update-based MIA directly leverages exchanged information to develop an attack model that distinguishes between members and non-members.
According to the exchanged information, these attacks can be further divided into those based on
\textit{model gradient} and \textit{model parameter}.
The former directly exploits original gradients or gradient differences as the input of attack models, whereas the latter extends the shadow training method to FL scenarios or uses a specialized structure of the target model to infer membership information.

\subsubsection{Model Gradient-based MIAs}

MIAs of this type in FL treat model gradients as part of attack feature vectors \cite{NMS+19, LHL+20, GUS+21}, or compare the gradients between rounds \cite{MLS+19, JLN+23} to infer the membership status of a query example.

\begin{figure}[htbp]
	\centering
	\includegraphics[width=3.8in]{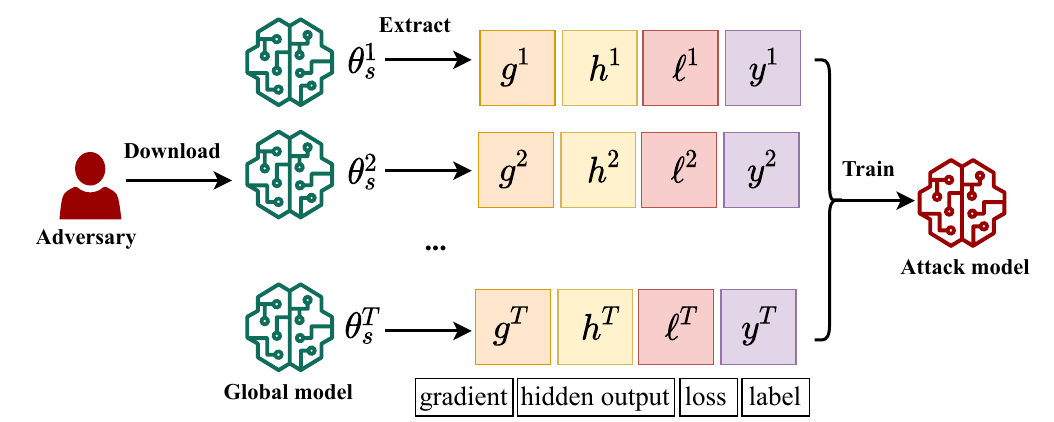}
	\caption{Overview of update-based MIAs by using original gradients and other intermediate features.}
	\label{fig:nasr}
\end{figure}

\textbf{Infer Membership via Original Gradient.}
This approach commonly regards the original gradients exchanged as one of the attack features and trains an attack model to discern the membership status.
Nasr et al.~\cite{NMS+19} develop a pioneering inference attack against the FedAvg algorithm~\cite{MBM+17}, which can infer leaked private information in the ML model by exploiting gradients and intermediate outputs in either an active or a passive manner.
A participant or server can gather original gradients, hidden layer outputs, loss values, and ground truth labels over multiple iterations and exploit them to construct an attack model in a passive strategy, as depicted in Fig.\ref{fig:nasr}.
Moreover, this work also develops an active attack strategy that causes victim participants to divulge more information through the \textit{gradient ascent algorithm}.
Specifically, for a query example $\textbf{x}$, an active attack performs gradient ascent as shown in Eq.(\ref{eq:ga}), where $\eta$ represents the local learning rate, and $L\left(\textbf{x}, \theta\right)$ denotes the loss on $\textbf{x}$.
\begin{equation}
	\small
	\label{eq:ga}
	\theta \gets \theta + \eta \frac{\partial L\left(\textbf{x}, \theta\right)} {\partial \theta}.
\end{equation}
Contrary to conventional SGD algorithms that reduce the loss of a training sample, an active attacker intentionally increases the loss on $\textbf{x}$ to widen the gap between the member and non-member data points.
If a query example is a member of the training data, the SGD algorithm decreases its loss after an honest participant updates the model. As such, a non-member record maintains a high loss due to the absence of additional modifications.
Furthermore, when the server is a malicious attacker, the attacker can isolate the victim participant to obtain a local view of the target model and expose more private information.
This study showcases how an adversary can improve attack performance in a white-box scenario. We conjecture the rationale behind it is that gradients confer an advantage for MIAs since they reveal more detailed information about the query example.
The gradients in a fully connected layer represent the inner products of the error from the next layer and the features at that layer~\cite{MLS+19}.

The work of Nasr et al.~\cite{NMS+19} has been extended to other applications \cite{LHL+20, GUS+21}.
A similar attack approach~\cite{LHL+20} is used to compare FL and coreset-based learning \cite{FDF+11, LHL+20} in terms of data privacy.
Specifically, the passive MIA method is used in FL, and a new attack strategy for coreset is established by K-means algorithm. The results show that FL is preferable over coreset for privacy protection with high accuracy. However, if a significant loss of model accuracy is tolerable, coreset can achieve privacy protection with less computation cost.
Gupta et al.~\cite{GUS+21} develop MIAs against deep regression that predicts brain age in the FL setting. This approach leverages gradients, activations, and predictions to conduct MIAs from the viewpoint of a participant.
The authors show that privacy breaches are more eminent when local data is skewed or non-IID among participants.

Although MIAs discussed above \cite{NMS+19, LHL+20, GUS+21} can be applied in various networks and scenarios, they have limitations in terms of assumptions and efficiency.
For instance, they require access to partial member data from victim clients, which may be challenging in scenarios with strict privacy constraints (e.g., medical area). Additionally, it is time-consuming to construct an attack model as it requires a significant number of feature vectors from various iterations.

\textbf{Infer Membership via Gradient Difference.}
To address the limitations of using original gradient, it is possible to directly infer the membership information by evaluating the gradient changes between consecutive iterations, given the distinction between members and non-members in the gradient distribution.
Melis et al.~\cite{MLS+19} introduce a novel MIA based on gradient difference in which the embedding layer leaks the membership status of training samples.
The core idea is that non-zero gradients in the embedding layer disclose which samples are trained in a batch.
Specifically, an honest-but-curious participant collects consecutive snapshots of the global model in  the $\left(t-1\right)$-th and $t$-th rounds, and calculates the difference between them as $\Delta \theta^{t}=\theta_{s}^{t}-\theta_{s}^{t-1}=\sum{}_{c}\Delta\theta^{t}_{c}$.
The model update contributed by other participants can be represented as $\Delta \theta^{t}-\Delta \theta_{adv}^{t}$, where $\Delta \theta_{adv}^{t}$ denotes the update of the adversary in the $t$-th iteration.
The proposed technique effectively detects the presence of location or text samples with a small batch size in a two-party FL setting. However, as the batch size increases, the false positive ratio significantly increases as it becomes more challenging to identify the exact training representation for a large set of word candidates in a bag-of-words format.
Moreover, this method is only applicable to neural networks with embedding layers and cannot be applied to deep learning models that use numeric data (e.g., tabular or image).

Li et al.~\cite{JLN+23} propose two passive MIAs called the gradient-diff attack and cosine attack. In these approaches, the adversary computes the gradient difference in consecutive rounds to deduce the membership status.
The gradient-diff attack is based on gradient orthogonality, through Eq.(\ref{eq:grad-diff}) to indicate the membership status of $\textbf{x}$, where $\Delta \theta_{c}^{t+1}$ denotes the local model update of client $c$ in the $\left(t+1\right)$-th iteration. If Eq.(\ref{eq:grad-diff}) holds, $\textbf{x}$ is a training record in the private set $D_c$.
\begin{equation}
	\small
	\label{eq:grad-diff}
	\left \| \Delta \theta_{c}^{t+1}  \right \| _{2}^{2} -
	\left \| \Delta \theta_{c}^{t+1} - {\textstyle \sum_{y\in Y} \nabla_{\theta} \ell \left(
			F\left(\textbf{x};\theta_{s}^{t}\right), y \right) }_{2}^{2} \right \| > 0.
\end{equation}
Moreover, the authors find a clear disparity between the distributions of the cosine similarity for member instances and non-member instances. Although both distributions resemble Gaussian distributions, their varying averages reveal distinct membership characteristics.
As such, they develop a cosine attack to deduce membership information by measuring the angle between these two vectors:
\begin{equation}
	\small
	\label{eq:cosim}
	\begin{aligned}
		\sum_{y\in Y} {\rm sgn} \left( {\rm cosim} \left( \nabla_{\theta} \ell \left(
		F\left(\textbf{x};\theta_{s}^{t} \right), y \right), \Delta \theta_{c}^{t+1}\right)\ge \gamma\right),
	\end{aligned}
\end{equation}
where $\rm sgn \left(\cdot\right)$ is an indicator function, $\rm cosim \left(\cdot, \cdot\right)$ denotes the cosine similarity, and $\gamma$ is a preset threshold based on non-member data.
Their MIAs are still effective and robust under privacy protection mechanisms \cite{AMC+16, BJZ+18}. 
Zhu et al.~\cite{ZGL+24} explore this idea further by extending it across multiple communication rounds and integrating models from non-target clients, boosting attack effectiveness significantly.

\subsubsection{Single Model-based MIAs}

MIAs in this category extract information about membership by utilizing either a historical global model or a local model in the FL training. Most of these attacks adapt the shadow training method from CL for conducting MIAs~\cite{PAM+20, YWY+23}, often enhanced through data augmentation~\cite{ZJZ+20, LZL+22}. Additionally, manipulating a specific structure within the target model can also facilitate membership inference in FL~\cite{PGR+22, JHP+22}.

\textbf{Infer Membership via Shadow Training.}
Inspired by the shadow training technique in CL, this approach considers the global or local model as the target model. However, in contrast to the CL scenario, it can be implemented more easily as any participant can naturally access the target model and auxiliary data.
Pustozerova et al.~\cite{PAM+20} develop an MIA for a sequential FL framework that exposes an individual local model to the subsequent participant immediately after training on private datasets, allowing others to infer information from the received model.
Assuming the presence of an auxiliary dataset, an attacker (e.g., one of the participants) builds shadow models, collects attack features, and develops an attack model to launch an inference attack on the victim model.
Luqman et al.~\cite{LAC+23} investigate CL-related MIAs based on shadow training in the peer-to-peer FL setting and reveal that membership leakage intensifies when colluding adversaries are involved. 
FD-Leaks, an MIA proposed by \cite{YZZ+23}, is tailored for federated distillation learning that involves clients exchanging model outputs on a public dataset.
Unlike the shadow training approach in CL, this method treats the attacker's model as the shadow model and avoids retraining.

\begin{figure}[htbp]
	\centering
	\includegraphics[width=4.0in]{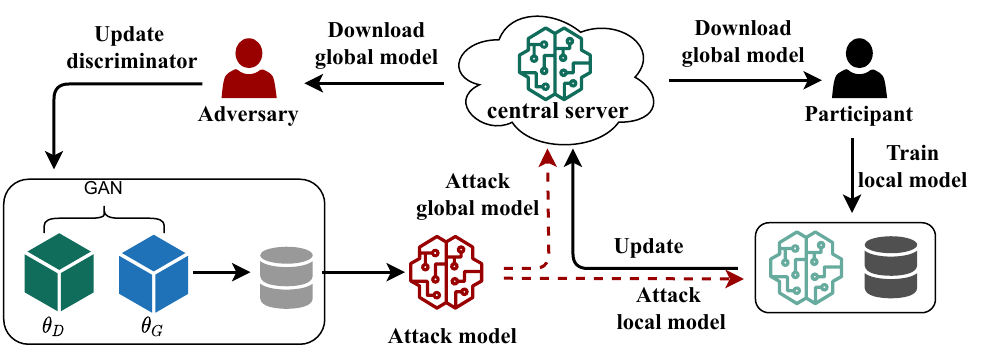}
	\caption{Overview of updated-based MIAs enhanced by data augmentation.}
	\label{fig:gan-data}
\end{figure}
Shadow training approaches can be enhanced by focusing on the construction of both shadow datasets and attack datasets~\cite{ZJZ+20, LZL+22, CJZ+20, ZYC+21, BSR+24}.
Data augmentation is implemented to boost the shadow dataset in the FL setting. As illustrated in Fig.\ref{fig:gan-data}, an attacker uses a shared global model as the discriminator of a GAN to generate diverse data and update it during the learning process. After generating sufficient attack data, the attacker trains a binary attack model by shadow training approach.
Zhang et al.~\cite{ZJZ+20} first employ GANs to enhance the MIAs launched by insiders.
Subsequently, Liu et al.~\cite{LZL+22} develop an MIA in which an eavesdropper behaves as a regular participant but has no knowledge of the private training dataset and attempts to attack the global model in FL.
Assuming that clients' labels are non-overlapping among different participants, source-level MIAs can be launched by comparing the query example's prediction result and label distribution among participants \cite{CJZ+20, ZYC+21}. 
In terms of attack data construction, MIA-BAD~\cite{BSR+24} improves shadow training methods by introducing a batch-wise attack dataset, inspired by the ensembling phenomenon~\cite{DTG+00}.

As for attacks beyond classification tasks, Yuan et al.~\cite{YWY+23} investigate an MIA that utilizes the shadow training approach against federated recommender systems (FedRecs), in which an honest-but-curious server aims to determine items interacted by a user based on uploaded parameters.
Attacks on classification tasks cannot be applied to FedRecs because most attacks aim to infer the existence of a query example, which is ineffective for FedRecs as only positive items (i.e., items interacted with by a user) disclose private information.
Specifically, the attack process involves training a shadow recommender model by randomly assigning ratings to the relevant items, then iteratively selecting the closest items, and finally retraining the shadow model until it reaches the preset number of guessed items.
This attack efficiently infers user interaction information for Fed-NCF and Fed-LightGCN frameworks~\cite{HXL+17, HXD+20}. Additionally, the authors also observe that membership information leakage could increase with more auxiliary knowledge, such as popular items.

\textbf{Infer Membership via Structure Modifying.}
In such attacks, a malicious server may actively manipulate model structures to infer membership information.
Pichler et al.~\cite{PGR+22} study an active attack by modifying the network of the client model. This approach uses the rectified linear unit (ReLU) property that the derivative of an output with respect to the parameters is zero when the output is negative.
Specifically, a malicious server embeds a network module equipped with ReLU activations into the target model, followed by configuring its parameters to enable activation by training members. When encountering an unseen query example, the parameters within the architecture remain unaltered. Consequently, the attacker determines the membership status by comparing the parameter changes with a predetermined threshold.
A similar idea is also explored in~\cite{JHP+22}, where a malicious server carefully crafts and embeds malicious parameters into a specific neuron, which a member sample can only activate.
The dishonest server can infer membership details by examining the neuron's gradient during the learning process. These attacks are simple yet effective, with the malicious strategy applicable to networks utilizing ReLU activations.

\begin{table*}[htbp]
	\setlength\tabcolsep{1.0pt}
	\footnotesize
	\centering
	\caption{
			Summary of update-based membership inference attacks on federated learning.
		}
	\label{tab:updateattack}
	\begin{tabular}{llllll}
		\toprule
		Year & Reference & Task & Technique & Comparison & Summary \\
		\hline 
		\hline
		2019 & Nasr et al.~\cite{NMS+19} & Classification & Intermediate output &  \cite{SRM+17} & \multirow{3}{*}{\makecell[l]{ Use original gradients \\ $\ominus$ Require large feature vectors \\ $\ominus$ Require auxiliary member data}} \\
		2021 & Gupta et al.~\cite{GUS+21} & Regression & Intermediate output & - & ~ \\
		2020 & Lu et al.~\cite{LHL+20} & Classification & Intermediate output & - & ~ \\
		\hline
		2019 & Melis et al.~\cite{MLS+19}  & Embedding & Non-zero gradient & - & \multirow{3}{*}{ \makecell[l]{Use gradient differences \\  $\oplus$ Require fewer feature vectors \\ $\oplus$ Low computation } } \\
		2023 & Li et al.~\cite{JLN+23} & Classification & Gradient orthogonality & \cite{JBW+21} & ~\\
		2024 & Zhu et al.~\cite{ZGL+24} & Classification  & Gradient similarity & \cite{YSG+18, NMS+19, JLN+23} & \\
		\hline
		2022 & Liu et al.~\cite{LZL+22} & Classification & Data reconstruction & - & \multirow{10}{*}{\makecell[l]{Extend shadow training to FL \\ $\oplus$ Launched by any role \\  $\oplus$ Enhanced by data augmentation \\ $\ominus$ Underexplored information}} \\ 
		2022 & Pustozerova et al.~\cite{PAM+20} & Classification & Sequential update & Random guessing & \\ 
		2020 & Zhang et al.~\cite{ZJZ+20} & Classification & Data augmentation & Random guessing & \\
		2020 & Chen et al.~\cite{CJZ+20} & Classification & Non-overlapping label & \cite{NMS+19} & \\
		2021 & Zhao et al.~\cite{ZYC+21} & Classification & Non-overlapping label & - & \\
		2023 & Luqman et al.~\cite{LAC+23} & Classification & Peer-to-peer update & - & \\
		2024 & Banerjee et al.~\cite{BSR+24} & Classification & Batch-wise feature & \cite{SRM+17} & \\
		2018 & Truex et al.~\cite{TSL+18} &  Classification & Decision boundary & Random guessing &  \\
		2023 & Yuan et al.~\cite{YWY+23} & Recommendation & Embedding relevance & \makecell[l]{K-means \\ Randow guessing} & \\
		\hline
		2022 & Pichler et al.~\cite{PGR+22} & Classification & ReLU activation & - & \multirow{3}{*}{\makecell[l]{Modify the target model \\ $\oplus$ Low computation \\ $\ominus$ Require specialized networks}}\\
		2022 & Nguyen et al.~\cite{JHP+22} & Classification & Poisoned neuron & - & ~\\
		& & & & & \\
		\bottomrule
		\end{tabular} 
		\begin{tablenotes}
			\item[1] $\oplus$: Advantages; $\ominus$: Disadvantages
		\end{tablenotes}
\end{table*}


To summary, it is intuitive to exploit original gradients to build an inference model. However, high computational and memory resources are required when collecting all intermediate outputs of target models.
In addition, these original gradient-based MIAs require an adversary to access partial member data to build a high-quality attack model \cite{NMS+19, LHL+20, GUS+21}.
In contrast, gradient difference techniques are proposed to tackle these challenges efficiently and practically.
Additionally, attacks based on model parameters mainly treat one of the snapshots as the target model and conduct MIAs to infer the private information of FL clients.
Given the white-box scenario in FL, the shadow training approach is practically implemented and boosted with the help of data augmentation via generative models. 
On the other hand, the structure modifying approach \cite{PGR+22, JHP+22} is limited to a malicious central server and linear layers, unsuitable for heterogeneous FL settings and complex networks.
We present additional details of trend-based attacks in Table~\ref{tab:updateattack}, including the year of application, the application domain, the specific methods employed, attacks used for comparison with the proposed approach, and a summary of each kind for readers' reference.

\subsection{Trend-based Membership Inference Attacks}

\begin{figure}[htbp]
	\centering
	\includegraphics[width=3.6in]{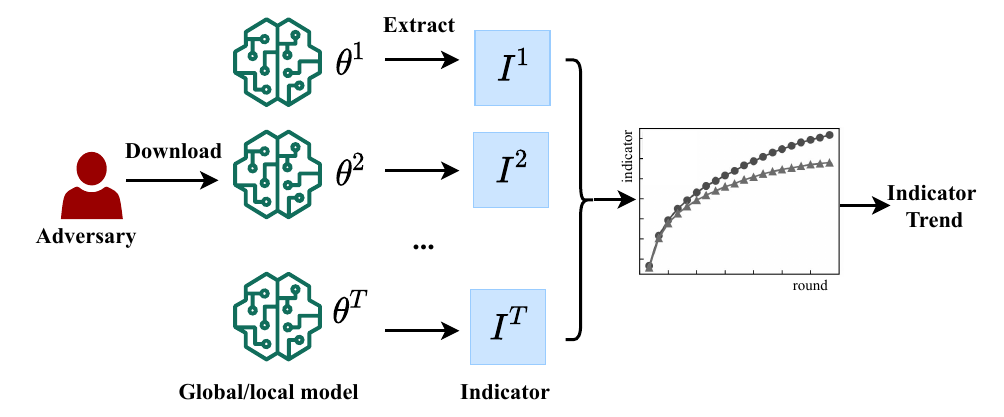}
	\caption{Overview of trend-based MIAs.}
	\label{fig:trend}
\end{figure}

Trend-based MIAs examine the evolution of an indicator associated with membership status to determine whether a record is a member during the learning process.
Such MIAs typically collect the historical models, extract the indicator information, and make decisions by comparing the indicator distributions between members and non-members, as shown in Fig.\ref{fig:trend}.
According to the indicator knowledge used, trend-based attacks can be categorized into those based on \textit{prediction score} and \textit{prediction loss}.

\subsubsection{Model Output-based MIAs}
For an ML model, the prediction score (a.k.a, confidence) and predication loss in the private data of participants often increases faster than that observed from testing data as the model converges, partially due to overfitting and memorization.
Inspired by this phenomenon, the difference in model output changes between training and test data across iterations is used when inferring membership information \cite{ZOX+21, GYB+22, ZYB+23}.

\textbf{Infer Membership via Prediction Trajectory.}
This approach uses the trend in confidence predicted by multiple models to differentiate between member and non-member samples.
A passive MIA proposed by \cite{ZOX+21} exploits a sequence of prediction scores generated by a local model to deduce the membership status.
Specifically, to infer the privacy of a specific client $c$, a curious eavesdropper gathers many versions of local models exchanged between $c$ and the server, and then calculates the prediction probability of the ground truth label of a query example $\textbf{x}$, namely,
$\left\{F_y \left(\textbf{x};\theta_{c}^{t}\right)\right\}_{t=1}^{T}$,
where $T$ is the number of rounds.
Then, the adversary constructs a fully convolutional attack model to classify these series and learn the difference between the training and test data.
This approach simplifies the input requirements of the method proposed by \cite{NMS+19}, as the input vector of attack models only consists of a single number in a round, and thus, a comparative attack performance can be achieved with lower computational resources and memory.

Similarly, Gu et al.~\cite{GYB+22} propose confidence-series-based MIAs (CS-MIAs) that use advanced confidence metrics and active attack strategies.
Different from~\cite{ZOX+21}, CS-MIAs use a novel confidence metric called {\it modified prediction entropy}~\cite{SLM+21} to calculate the confidence series.
In addition, CS-MIAs allow a global adversary (i.e., the server) to extract more private information by active participation and selection.
The global adversary fine-tunes the global model using auxiliary data and submits updates similar to a regular participant in each iteration to address the problem of lack of training data when building the attack model.
This process makes the confidence scores on the shadow training data similar to those on the target training data. Furthermore, the adversary deliberately selects the target client in each round instead of random selection to force the victim participant to leak more private information.
Experimental results demonstrate that CS-MIAs outperform existing state-of-the-art black-box MIAs \cite{SLM+21, SAZ+18, NMS+19}.

Zhang et al.~\cite{ZYB+23} explore the poisoning MIA (PMIA) in FL, where poisoning attacks are employed to enhance the effectiveness of MIAs.
This approach injects malicious gradients to improve ML models by triggering victim clients to repair manipulations, leading to disclosing membership information to the attacker.
During the training phase, this approach formulates an optimization problem to maximize the loss of victim training data while evading detection from Byzantine-robust aggregation mechanisms by introducing a ``cover'' dataset to conceal the adversary's update. Subsequently, membership information is inferred using prediction correctness or trend analysis.
It is noteworthy that the prediction trends behind PMIA differ from the aforementioned works. Specifically, if a client observes an incorrect prediction on a training sample after an attack round, the subsequent training rounds are expected to reach a correct prediction.

Rather than employing prediction series directly, Liu et al.~\cite{LGT+23} presented a novel attack technique based on the temporal evolution of adversarial robustness when an adversary's access is restricted solely to prediction labels. Their approach stems from the conspicuous disparities observed in the convergence patterns of adversarial robustness between training and test data.

\textbf{Infer Membership via Loss Trajectory.}
This approach typically examines the change in the loss of samples throughout the training process to infer the membership status.
Suri et al.~\cite{SAK+22} propose a subject-level MIA that aims to infer the privacy of a particular individual’s (a.k.a, subject’s) data in the cross-silo FL setting.
Subject-level privacy is equivalent to source-level privacy in the cross-device FL setting, in which FL has a one-to-one mapping between data subjects (each individual owning a participating subject). This equivalence no longer works in the cross-silo setting when an individual’s data is spread across several federation users or organizations \cite{SAK+22, MVJ+22}.
By assuming an auxiliary dataset about inferred subject distribution, the authors develop a loss-across-round attack to infer the subject membership.
The loss-across-round attack from a participant or server exploits the change in the loss values as the training rounds progress to deduce the subject membership status. To this end, the attacker records loss values of the subject’s dataset $D_s$ across each training round $t$, counts the number of training rounds where the loss decreases, and finally compares the final value $\mathcal{L}$ with a threshold by Eq.(\ref{eq:loss-ac}).
In addition to utilizing multi-round model updates from the target client, Zhu et al.~\cite{ZGL+24} enhance the effectiveness of MIAs by incorporating models from non-target clients, particularly in scenarios with homogeneous data distributions.
\begin{equation}
	\label{eq:loss-ac}
	\begin{aligned}
		\mathcal{L}_t=\sum_{\left(\textbf{x},y\right)\in D_s} \ell_t \left(\textbf{x},y\right)  \\
		\mathcal{L} = \sum_{t=1}^{r}\mathbb{I}\left[\mathcal{L}_t < \mathcal{L}_{t-1}\right]
	\end{aligned}
\end{equation}

Hue et al.~\cite{HHS+21} targets source-level privacy and leverages the prediction loss to determine the source of a training sample from a server perspective. This is based on the fact that the probability of an example being a member depends on its loss~\cite{SAM+19}.
SIA allows an honest-but-curious server to launch a source-level attack in the FL setting by comparing the loss across local models.
The source of that example is the participant whose local model yields the smallest loss. The success of SIA relies on the generalization of local models, which highlights the privacy risks arising from the non-IID phenomenon in FL.
Furthermore, SIA provides an alternative approach for an FL server that conducts inference attacks against a specific participant, that is, it can leverage out-of-the-box MIAs on CL, such as metric-based attacks \cite{SLM+21, SLS+19-1, YSG+18}, to easily breach the privacy of participants' local data.

\subsubsection{Model Parameter-based MIAs}

The optimization of an ML model aims to reduce the difference between its output and the true output on the training set, resulting in adjustments to the model's parameters. MIAs based on model parameter exploit these changes by identifying distinct patterns in how the model behaves for member versus non-member samples.

\textbf{Infer Membership via Bias Trajectory.}
Zhang et al.~\cite{ZLL+23} are the first to leverage changes in model bias to determine membership in federated learning (FL). Their approach was motivated by the observation that non-member samples induce significant changes in model parameters, resulting in a more pronounced shift in bias. 
This work specifically examines the bias values of the final layer and incorporates feature amplification through an exponential function. 
By gathering the model over multiple epochs, it calculates the bias changes and achieves membership inference. Empirical evidence suggests that these bias differences incur minimal overhead while providing effective features for MIAs.


To conclude, trend-based attacks exploit the trajectory of model output or parameter to deduce membership status and exhibit the following advantages. 
First, instead of collecting a large feature vector like \cite{NMS+19, LHL+20, GUS+21}, they gather only a single value in a round.
Moreover, these approaches can make decisions about membership information by comparing the indicator with a preset threshold or observing the trend direction of the indicator without developing an attack classifier model. These advantages lower the computation and memory resources cost and the implementation complexity.
We present additional details of trend-based attacks in Table~\ref{tab:trendattack}, including the year of application, the application domain, the specific methods employed, attacks used for comparison with the proposed approach, and a summary for each kind.

\begin{table*}[htbp]
	\setlength\tabcolsep{1.0pt}
	\footnotesize
	\centering
	\caption{Summary of trend-based membership inference attacks on federated learning.}
	\label{tab:trendattack}
	\begin{tabular}{llllll}
		\toprule
		Year & Reference & Task & Technique & Comparison & Summary \\
		\hline 
		\hline
		2021 & Zari et al.~\cite{ZOX+21} & Classification &  Predication sequence & \cite{NMS+19} & \multirow{4}{*}{\makecell[l]{Use prediction trajectory \\ $\oplus$ Require fewer feature vectors \\ $\ominus$ Require ground  truth labels}} \\
		2022 & Gu et al.~\cite{GYB+22} & Classification& Modified predication sequence & \cite{NMS+19} & \\
		2023 & Zhang et al.~\cite{ZYB+23} & Classification & Poisoning gradient & \cite{MLS+19, NMS+19} & \\
		2023 & Liu et al.~\cite{LGT+23} & Classification & Adversarial robustness & \cite{NMS+19, CCC+21} & \\
		\hline
		2021 & Hu et al.~\cite{HHS+21} & Classification & Loss comparison & Random guessing & \multirow{3}{*}{\makecell[l]{Use loss trajectory \\ $\oplus$ Require fewer feature vectors \\ $\ominus$ Require ground  truth labels}} \\
		2022 & Suri et al.~\cite{SAK+22} & Classification  & Multi-round loss & Random guessing & \\
		2024 & Zhu et al.~\cite{ZGL+24} & Classification & Multi-round-client loss & \cite{YSG+18, NMS+19, JLN+23} & \\
		\hline 
		2023 & Zhang et al.~\cite{ZLL+23} & Classification & Multi-round bias & \cite{NMS+19, LYZ+22, SLS+19-1} & \makecell[l]{Use parameter changes \\ $\oplus$ Require fewer feature vectors \\ $\oplus$ Without label requirement} \\
		\bottomrule
		\end{tabular} 
		\begin{tablenotes}
			\item[1] $\oplus$: Advantages; $\ominus$: Disadvantages
		\end{tablenotes}
\end{table*}

\begin{table*}[htb]
	\footnotesize
	\centering
	\caption{A comparison of attack approaches in centralized and federated learning.}	\label{tab:comp-attack}
	\begin{threeparttable}
		\begin{tabular}{c|c|c|c|c|c}
			\toprule
			\bf{Setting} & \bf{Type} & \bf{Approach} & \bf{Phase} & \bf{Attack Feature} & \bf{Unique}\tnote{1}\\
			\hline
			\hline
			\multirow{5}{*}{CL} & Classified-based  & Shadow training & \multirow{5}{*}{Inference} & \makecell{Model output \\ Intermediate output}  & \Circle \\
			\cline{2-3} \cline{5-6}
			
			~ & \multirow{4}{*}{Metric-based} & Prediction correctness & ~ & Model output & \CIRCLE \\
			\cline{3-3} \cline{5-6}
			~ & ~ & Prediction loss & ~ & Loss of model output & \LEFTcircle \\
			\cline{3-3} \cline{5-6}
			~ & ~ & Prediction confidence & ~ & Model output & \LEFTcircle \\
			\cline{3-3} \cline{5-6}
			~ & ~ & Prediction entropy & ~ & Model output & \CIRCLE \\
			\hline
			\hline
			
			\multirow{9}{*}{FL} & \multirow{5}{*}{Update-based}  & Original gradient & Aggregation & \makecell{Model output  \\ Gradient \\ Intermediate output} & \CIRCLE \\
			\cline{3-6} 
			~ & ~ & Gradient difference & \makecell{Aggregation \\ Communication} & \makecell{Gradients within \\ consecutive iterations }& \CIRCLE \\
			\cline{3-6} 
			~ & ~ & Shadow training & Local training & Global model & \Circle \\
			\cline{3-6} 
			~ & ~ & Structure modifying & Local  training & Global model & \CIRCLE \\
			\cline{2-6} 
			
			~ & \multirow{3}{*}{Trend-based} & \makecell{Prediction trajectory \\ Loss trajectory} & Local training & \makecell{Model output \\ within several iterations}   & \LEFTcircle \\
			\cline{3-6} 
			~ & ~ & Bias trajectory & Local training & \makecell{Model bias \\ within several iterations} & \CIRCLE \\			
			
			\bottomrule
		\end{tabular}
		\begin{tablenotes}
			\item[1] \CIRCLE: unique attack \LEFTcircle: partially unique attack \Circle: common attack
		\end{tablenotes}
	\end{threeparttable}
\end{table*}

\subsection{Comparison to Attacks in Centralized Learning}
In this section, we examine the existing update-based and trend-based MIAs in the context of FL.
Update-based approaches focus on extracting membership information from model updates exchanged between clients and the server, including model gradients and historical models throughout the federated learning process. In contrast, trend-based approaches analyze the evolving patterns of the training data to conduct MIAs, capitalizing on the observation that member samples exhibit distinct behaviors compared to non-member samples from an indicator perspective as the learning process progresses.

For a comprehensive understanding of MIAs in the FL context, we compare them with attacks relevant to CL, encompassing both classifier-based and metric-based methods mentioned in Section~\ref{sec:mia}. 
Table~\ref{tab:comp-attack} highlights key differences between FL-related MIAs and those in CL, focusing on two main aspects: the phase in the model's lifecycle when the attack occurs, and the adversarial knowledge leveraged to infer membership. 
These differences give rise to innovative attack methodologies in the FL setting. 
With access to the internal details of the target model, an attacker can leverage additional adversarial knowledge to enhance the effectiveness of inference attacks, employing the original gradient or gradient difference methods. However, these techniques are generally not applicable to most black-box MIAs in CL.
Moreover, in FL, adversaries often have access to historical versions of the target model, enabling them to track the trajectory of query examples over time. 
This can greatly increase the risk of membership information leakage and supports trend-based attacks in FL, whereas such information is typically less available in CL scenarios.

Additionally, 
source-level MIAs are distinct in the FL setting because of its collaborative training strategy.
From the perspective of the central server, this is intuitively equivalent to using record-level MIAs to sequentially infer the membership status of each local model~\cite{ZJZ+20, CJZ+20}.
However, for a more effective and efficient attack, it is essential to consider the local models from all clients simultaneously. 
Hu et al.~\cite{HHS+21} compare the loss values across clients and identify the client with the lowest loss as having the source dataset.
Zhang et al.~\cite{ZLL+23} leverage the bias differences in the final layer and analyze multiple model epochs to identify the member source. They assign the data to the participant exhibiting the smallest change in bias.
Current MIA research primarily emphasizes record-level privacy risks, as the membership leakage of an individual sample is central to private information exposure. Additionally, most record-level attacks can be extended to source-level attacks~\cite{ZLL+23}.

 \section{Membership Inference Defenses in Federated Learning}
\label{sec:defense}

Various defenses have been proposed to alleviate the growing concerns regarding private information leaked through MIAs in the FL setting.
In this section, we review existing defense strategies and divide them into four categories, namely, partial sharing, secure aggregation, noise perturbation and anomaly detection.
Each category is classified further based on specific approaches, as illustrated in Table~\ref{tab:defense1}.

\begin{table*}[htbp]
	\setlength\tabcolsep{0.5pt}
	\footnotesize
	\centering
	\caption{Taxonomy of membership inference defenses in federated learning.}
	\label{tab:defense1}
	\begin{tabular}{c|c|c|c}
		\toprule
		\textbf{Type} & \textbf{Approach} &  {\centering \textbf{Advantages}} & {\centering \textbf{Disadvantages}}  \\
		\hline
		\hline
		
		\multirowcell{2}{\centering Partial sharing} & Gradient compression 
		& \multirowcell{1}{\makecell[l]{
				Slight utility loss \\
				Low computation cost \quad \quad \quad \quad \quad \quad
		}}
		& \multirowcell{1.5}{\makecell{
				Limited mitigation effect \quad \quad \quad 
		}}  \\
		~ & Weight pruning  ~ & ~ & \\
		\hline
		\multirowcell{3}{\centering Secure aggregation} & Secure multi-party computation 
		& \multirowcell{1}{\makecell[l]{			
				Lossless model utility \\
				Without a trusted server \\
				Protection from the central server \, \\
		}}
		& \multirowcell{1.5}{\makecell[l]{
				High computation cost \\
				Vulnerable to malicious clients
		}}  \\
	
		~ & \multirowcell{2}{Homomorphic encryption}  ~ & ~ & \\
		~ & ~ & ~ & ~ \\
		\hline
		\multirowcell{2}{\centering Noise perturbation} & Differential privacy 
		& \multirowcell{1}{\makecell[l]{
				Strong privacy guarantee \\
				Protection from a server and clients
		}}
		& \multirowcell{1.5}{\makecell[l]{
				High model utility loss \quad \quad  \quad \quad
		}}  \\
		~ & Random perturbation  ~ & ~ & \\
		\hline
		Anomaly detection & Misbehaving identification & Defend against poisoning MIAs \quad \, \quad   & Fail for passive attacks \quad \quad  \quad \quad \\
		\bottomrule
	\end{tabular}
\end{table*}

\subsection{Partial Sharing}
Partial sharing aims to reduce the effectiveness of inference attacks by suppressing certain updates exchanged during the FL process.
Since the internal state of FL models is susceptible to MIA attacks through sharing parameters or gradients, one straightforward approach is to limit the information available to adversaries by reducing the amount of data shared~\cite{NNJ+22}.
Defenses based on this category can be further divided into \textit{gradient compression} and \textit{weight pruning}.

\subsubsection{Gradient Compression}
This defense strategy does not upload all gradient tensors but transmits only a few of them via selective strategies, such as by choosing the top-$K$ largest values \cite{SRZ+15, AAH+17, SSU+18} or those exceeding a certain threshold~\cite{DAB+20}.
The intuition  is that although a global model depends on local updates to learn knowledge from training data, not all updates equally contribute to model parameters.
In addition to privacy protection, this strategy can also decrease the transferred data volume and save communication costs during the FL process.

A partial sharing algorithm, namely, distributed selective SGD (DSSGD)~\cite{SRZ+15}, uses only a tiny fraction (i.e., 1\%) of gradients shared per-participant to achieve better performance than CL.
DSSGD allows an FL participant to download the latest global model, compute the top-$K$ largest gradients, and upload them to the server.
Melis et al.~\cite{MLS+19} explore the use of DSSGD to defend MIAs on a two-party sentiment classifier and show that the attack accuracy decreases from 0.93 AUC to 0.84 AUC when only 10\% of the updates are shared during FL.

In addition to intuitive selection, researchers have also developed advanced techniques to share fewer gradients while maintaining the model performance.
For example, deep gradient compression (DGC)~\cite{LYH+17} incorporates momentum correction and local gradient clipping after gradient sparsification to preserve the model performance. Moreover, warm-up training is introduced to address the staleness issue~\cite{DWZ+19} in the learning process. By using only 0.1\% of the gradient exchange in distributed SGD, this approach achieves a high compression magnitude of 270-600x.
This work~\cite{YTH+18} proposes a novel gradient compression technique that delays the transmission of ambiguously estimated gradient elements whose amplitude is small than their variance over the data points. These methods can drastically compress the exchanged gradients while maintaining the original model's accuracy.

Instead of directly exchanging the gradient values, 
signSGD~\cite{BJZ+18} exchanges the sign of the gradient through majority vote aggregation. This technique enables convergence in large-scale and mini-batch datasets with theoretical and empirical evidence.
Recent work~\cite{JLN+23} has shown that signSGD is an effective countermeasure against MIAs, and the validation accuracy of the target model decreases by less than 1\%. Moreover, signSGD provides a powerful defense against label inference attacks~\cite{FCZ+22} and Byzantine attacks, even in the case of up to 50\% of adversarial workers.

An extreme solution for compression is to hide all raw gradient values and directions~\cite{CHS+19}. In essence, this approach exchanges the predictions of local models and obtains a robust mean estimation on all predictions, thereby mitigating MIAs and poisoning attacks~\cite{NMS+19}.
Intuitively, the release of only predictions is safer than the release of gradients, as high-dimensional gradients encode more information pertaining to the local data.
Since this approach essentially transforms the attack setting from white-box to black-box, off-the-box CL defenses can be integrated to enhance the defense performance.
Moreover, this approach can be applied in heterogeneous FL, as opposed to aforementioned approaches that are only suitable for homogeneous FL.

\subsubsection{Weight Pruning}
Since FL frameworks facilitate the exchange of model parameters among participants, weight pruning transmits only a few model parameters rather than all of them from each participant.
Inspired by weight pruning techniques in ML, researchers~\cite{SDG+22} develop an approach to pruning parameters in the global model to guard against MIAs. A sparsified model containing fewer than 5\% of the original model parameters can achieve comparable performance to that of the original model while more effectively mitigating MIAs.
However, balancing the sparsity and accuracy of pruned neural networks may require model retraining, and the reuse of training samples can increase the risk of privacy violations through memorization~\cite {YXZ+22}.

In a nutshell, partial sharing defense aims to exchange incomplete model updates to mitigate the threat of MIAs and lower communication costs without significant utility loss in FL. In addition, extreme compression methods can defend against other privacy and security threats, such as model inversion attacks \cite{LYH+17, YTH+18} and Byzantine attacks~\cite{BJZ+18}.
However, such approaches only provide limited protection of membership privacy \cite{AYH+17, MLS+19}, as they often have to exchange gradients or parameters that contain essential information about the training data to maintain the original performance.

\subsection{Secure Aggregation}
Secure aggregation typically relies on cryptographic techniques to prevent information leakage from potential adversaries.
The underlying idea is to disclose the encrypted model updates instead of the clear ones throughout the learning process.
This privacy-preserving approach can be divided into
\textit{secure multi-party computation (SMC)} and \textit{homomorphic encryption (HE)}.

\subsubsection{Secure Multi-party Computation}
SMC is a conventional technique to safeguard input when multiple participants collectively compute a specific output \cite{YAC+82, ZJL+21}. An effective SMC protocol must fulfill two essential requirements \cite{LRX+18, YXZ+21}: correctness, i.e., the outcome computed by the protocol must be accurate; and privacy, i.e., the information of no participant should be revealed to others.
SMC restricts the central aggregator from learning only the summation or average of the updates of clients and safeguards individual local updates from potential eavesdroppers or a curious server.

Active research has been conducted on SMC for the protection of sensitive information. For example, Mohassel et al.~\cite{MPZ+17} introduce a groundbreaking protocol for safeguarding privacy in ML, which enables the computation of non-linear activation functions, allows SGD optimization, and supports linear regression and neural networks.
Following this study, SecAgg~\cite{BKI+17} is proposed to protect FL from an honest-but-curious server. This protocol uses secret sharing and double-masking approaches to ensure that the server gains no knowledge beyond an aggregated model and cannot observe clear individual local models.
Similarly, Sayyad et al.~\cite{SS+20} generate secret shared entities to address privacy concerns for the data involved in deep learning.
However, SMC protocols suffer from high computational costs in real-world applications.
To address this problem, Fereidooni et al.~\cite{FHM+21} propose an efficient aggregation protocol named SAFELearn, which prevents client-side information leakage and limits the access of the central aggregator access to only the aggregated results of client updates. This approach has garnered significant attention because it does not rely on a trusted server, requires only two communication rounds, and allows clients to drop out.

While existing studies assert that SMC can safeguard participants' private information in FL, its practical and theoretical effectiveness as a defense mechanism remains limited~\cite{MLS+19, FHM+21, NKH+24}.
Previous research~\cite{GYB+22} shows that SecAgg~\cite{BKI+17} fails to guarantee data privacy against client-side MIAs, and \cite{GJH+21} points out that a malicious server can collude with other clients to compromise privacy, even with the use of SMC.
Additionally, this study\cite{NKH+24} presents a formal analysis of SecAgg against MIAs from the perspective of DP, arguing that its privacy guarantee depends on the model's dimensionality and the number of clients. It theoretically concludes that SecAgg provides weak privacy protection in FL, as the model size is usually much larger than the client pool.

To address these limitations, hybrid countermeasures have been developed to prevent information leakage from a curious participant that can access the global model.
Truex et al.~\cite{TSB+19} present a privacy-preserving approach that prevents intermediate messages and the final trained model from inference attacks. This approach combines SMC and differential privacy~\cite{DCM+06} to achieve privacy protection without sacrificing much accuracy. Each participant adds noise to their local model, which is then encrypted using the Paillier cryptosystem before being sent to the server.
By combining both defenses, this approach ensures data privacy with little utility loss.

\subsubsection{Homomorphic Encryption}
HE~\cite{RRL+78} is a practical technique to protect sensitive data by encrypting the uploaded parameters in FL. An HE cryptosystem \textbf{H} provides an operation $\star$ that satisfies $\textbf{H}\left(m_1\right) \star \textbf{H}\left(m_2\right) = \textbf{H}\left(m_1 \star m_2\right)$, where $m_1$ and $m_2$ are sets of plaintext. In essence, this technique allows certain operations to be performed in the ciphertext space without restoring the plaintext,  and thus, the performance of the training model will not be affected \cite{RRL+78, YXZ+21, LRX+18}.

Several privacy-preserving approaches for FL have been developed based on HE.
Phong et al.~\cite{AYH+17} exploit HE to construct a secure FL system, thereby alleviating the data leakage from semi-trusted third-party servers.
Notably, although this method can prevent the attacker from directly obtaining the local data of other parties, the attacker can still infer the distribution of private data~\cite{LZL+22}.
However, since HE involves high computational and communication overheads, researchers have attempted to simplify the aggregation scheme when protecting privacy in FL.
Bai et al.~\cite{BYF+21} combine HE and a selective parameter scheme to defend against MIAs and poisoning attacks from participants and the server.
Designed for a cross-silo FL scenario, BatchCrypt~\cite{ZCL+20} encodes a batch of quantized gradients into a long integer instead of treating full-precision gradients while incurring an accuracy drop of less than 1\%, helps accelerate the training process and significantly reduces the computation and communication costs associated with HE.
Ma et al.~\cite{MXS+22} propose a novel aggregation protocol using individual client model initialization and model updating to prevent eavesdroppers from inferring the local and global models, resulting in privacy enhancement and lowered computational costs in FL.

To summarize, secure aggregation utilizes encryption methods to reveal essential information to designated participants while safeguarding private data from being deduced~\cite{LYK+22}.
Furthermore, it can deal with the encrypted updates without affecting the training results of the model~\cite{ZCX+21}.
Nonetheless, cryptographic algorithms introduce computational and communication overhead. Additionally, a curious participant may compromise the privacy of a global model protected by SMC and HE~\cite{GYB+22}.

\subsection{Noise Perturbation}

Noise perturbation relies on the idea that the introduction of noises can hinder adversaries from discerning sensitive information during inference. These noises are incorporated into the target model in accordance with privacy requirements or optimization objectives.
This defense category contains both \textit{differential privacy (DP)} and \textit{random perturbation}.

\subsubsection{Differential Privacy}
DP \cite{DCM+06, DC+08, ZHH+20} is a lightweight privacy-preserving technique in which noise is added to sensitive data to offer strict privacy protection. It can be formally defined as follows:
	A randomized mechanism $\mathcal{M}: \textbf{X}^n \to \textbf{Y}$ satisfies $\left(\epsilon,\delta\right)$-DP for input $\textbf{X}^n$ and output $\textbf{Y}$,
	if for two neighboring subsets $D, D^\prime \in \textbf{X}^n$ that differ at most one element and for any $Y \in \textbf{Y}$, the following inequality holds:
	\begin{equation}
		\label{eq:dp}
		P\left[\mathcal{M}\left(D\right)\in Y\right] \le e^\epsilon P\left[\mathcal{M}\left(D^\prime\right)\in Y\right]+ \delta,
	\end{equation}
where $\epsilon$ refers to the privacy budget, and $\delta$ is the failure probability.
In essence, DP is a natural countermeasure to MIAs, as DP constrains the discrepancy between the presence and absence of a sample.
Extensive research based on theoretical and empirical perspectives has proven that DP can ensure data privacy against MIAs.

\textbf{(i) Effectiveness of DP from the Theoretical Perspective.}

Several studies have demonstrated the effectiveness of DP in thwarting MIAs and derived theoretical leakage upper bounds based on different assumptions and attack evaluation metrics \cite{YSG+18, EUM+19, BDE+21, SAM+19}.
Yeom et al.~\cite{YSG+18} draw inspiration from the correlation between DP and model generalization to establish a formal association between DP and MIAs.
The authors introduce a new metric, \textit{membership advantage}, which evaluates the attack performance by measuring the disparity between the true positive rate (TPR) and false positive rate (FPR), and demonstrate that the advantage of an attacker from a DP model is smaller than $e^{\epsilon}-1$.
Later, based on the proposition in \cite{HRR+13} that limits the TPR of any test to $\left(\epsilon,\delta\right)$-DP, Erlingsson et al.~\cite{EUM+19} derive a tighter bound for the membership advantage, that is, $1-e^{-\epsilon}\left(1-\delta\right)$.

The abovementioned bounds are derived under two restrictive assumptions: (1) the attacker derives membership information from a black-box model, and (2) the member and non-member instances adhere to the IID principle.
These assumptions do not always hold in FL settings, which involve white-box target models and non-IID data distributions. As such, researchers have attempted to extend them to realistic settings associated with FL scenarios.
Considering a white-box setting, Bernau et al.~\cite{BDE+21} establish the expected membership advantage for an all-powerful adversary that can access arbitrary background information, except for one identified record. The adversary is assumed to be able to obtain the neighborhood dataset and observe the gradient during model training, similar to an insider attacker in FL.
Motivated by data dependency in the training samples, other researchers~\cite{HTO+20} explore the protection strength of DP when statistical dependencies exist among instances.
According to a strong-privacy membership experiment \cite{NMSS+21}, the authors derive an upper bound for the membership advantage under $\left(\epsilon,\delta\right)$-DP as $\left(e^{\epsilon}-1+2\delta\right)/\left(e^{\epsilon}+1\right)$.
However, this bound about membership advantage can be very large in non-IID scenarios when there are statistical dependencies.

Different from theoretical bounds derived from membership advantage \cite{YSG+18, EUM+19, BDE+21}, Sablayrolles et al.~\cite{SAM+19} focus on the likelihood of an instance being a member and derive an upper bound of this probability under $\left(\epsilon,\delta\right)$-DP.
An ML model can be treated as a posterior distribution over parameters, and the probability of an instance being a member is bounded as $\lambda+\frac{\epsilon}{4}+\delta$, where $\lambda$ denotes the prior probability determined through random guessing.
Although these theoretical works indicate that DP can reduce the effectiveness of MIAs, they do not accurately reflect the risks of MIAs in practice, particularly for large $\epsilon$ \cite{YSG+18, HTO+20}.

\textbf{(ii) Effectiveness of DP from the Empirical Perspective.}

A substantial body of literature has presented empirical evidence for the effectiveness of DP, mainly by local DP (LDP) \cite{AMC+16, TSL+20} or central DP (CDP) \cite{MHB+18, GRC+17}, in mitigating the threat of MIAs in FL.
In LDP-based approaches, a participant chooses a DP mechanism $\mathcal{M}$ and adds noise locally to their data under privacy requirements to achieve record-level (individual data record) protection for private data.
By contrast, in CDP, noise is added to the aggregation model on the server, which renders the model outputs indistinguishable from adversaries and helps ensure client-level (users who participate in FL) privacy against MIAs.

\textit{\textbf{LDP-based Defenses}}.
Researchers have used LDP to conceal the effect of individual training examples within the private dataset of a client \cite{RMA+18, JBE+19, NNJ+22}.
DPSGD~\cite{AMC+16} is a typical algorithm to implement LDP. A moment accountant scheme is used to assign the privacy budget in the learning process.
Rahman et al.~\cite{RMA+18} use DPSGD as a defense against MIAs for an ML model. However, the authors highlight that realizing perfect protection requires a stringent privacy budget of $\epsilon \le 2$, which comes at the cost of reduced model utility.
Mohammad et al.~\cite{NNJ+22} find that LDP can effectively protect membership information and reduce the attack accuracy from 73\% to 52\%.
Hu et al.~\cite{HHS+21} observe that vanilla LDP is ineffective against their attacks based on the prediction loss in FL settings. When the privacy budget is smaller than 2, a defender is close to random guessing at the expense of significant model utility loss.

Some efforts have been made to address the trade-off between LDP effectiveness and degraded utility.
Relaxed versions can decrease the utility loss under the same privacy requirement, but they reveal more private information~\cite{JBE+19}.
A hybrid countermeasure~\cite{LZL+22} is developed by combining LDP with trust domain division.
The proposed approach exploits authentication based on certificates issued by a trusted certificate authority, restricts communication to certified clients, prevents potential eavesdroppers from inferring, and helps balance the model utility and privacy guarantee.



\textit{\textbf{CDP-based Defenses}}.
Studies on CDP offer client-level privacy protection by concealing the individual contributions of participants to FL \cite{MLS+19, AMM+21, NNJ+22, SAK+22}.
In this approach, the server clips the $l_2$ norm of each participant's update, aggregates the clipped updates, and then adds noise to the aggregated result to ensure privacy~\cite{MHB+18, GRC+17}.

Several studies have validated the effectiveness of CDP against MIAs in practice.
Mohammad et al.~\cite{NNJ+22} investigate the suitability of CDP in ensuring privacy protection and robustness against MIAs and backdoor attacks and demonstrate that CDP mechanisms could successfully prevent the white-box passive and active MIAs proposed in~\cite{NMS+19}.
The defense reduces attack accuracy from 75\% to 52\% with a utility drop of about 16\% on the CIFAR100 dataset.
By examining the performances of DP with various granularities~\cite{AMC+16, MVJ+22, MVJ+22}, Suri et al.~\cite{SAK+22} highlight that client-level privacy protection provides the best defense against membership information leakage, and LDP provided the least protection.
Despite its promising potential, the practical application of this approach faces challenges in some scenarios.
For example, Melis et al.~\cite{MLS+19} conduct empirical evaluations in a federated learning (FL) setting with fewer than 30 participants. They find that the global model fails to converge because the magnitude of the noise added is inversely proportional to the number of participants.

\subsubsection{Random Perturbation}
This method aims to protect private information by introducing carefully crafted noise under an optimization objective that minimizes the model utility loss and provides a privacy guarantee.

Well-crafted perturbations can be introduced into model updates while conducting FL training.
An accuracy-lossless noise perturbation method~\cite{YXF+22} can address the problem that DP offers solid theoretical guarantees but invariably impairs model accuracy \cite{HBA+17, MLS+19} by adding removable noise in the FL setting.
Inspired by the random sketching technique~\cite{ZMW+21} used to defend against property inference and model construction attacks, researchers~\cite{YXF+22} developed a technique to prevent malicious clients from accessing true global model parameters and local gradients through Hadamard products and linear outputs.
This approach reduces an attack accuracy of approximately 50\% while maintaining the learning accuracy. However, this approach applies only to networks that use ReLU as the activation function and cannot be extended to networks involving sigmoid and tanh functions. Additionally, it relies on a trustworthy server that will not attempt to infer private information from the recovered updates.
Besides, with reference to the MemGuard approach~\cite{JJS+19} against black-box attacks in the context of CL, Xie et al.~\cite{XYC+21} devise a noise addition technique to deceive adversaries into random guessing the membership status in the FL setting.

Another research line within this type involves perturbing the model's input data.
%
Yang et al.~\cite{YYY+23} introduce a client-level input perturbation called CIP to enhance the FL framework's data privacy by altering each client's local data distribution. 
Similarly, Lee et al.~\cite{LHK+21} design a digestive neural network for private data and provide anonymized training inputs to mitigate the risk of disclosing membership information.

In short, noise perturbation reduces the dependence between model performance and input data to maintain private information by adding noise.
Theoretical and empirical studies have demonstrated that DP schemes allow participants to maintain local data privacy within the FL framework. However, the effectiveness of DP often comes at the cost of significant utility loss, and it is challenging to select a suitable privacy budget to balance privacy and utility.
In contrast, random perturbation adds well-crafted noises restricted by optimization objectives, leading to a better utility-privacy trade-off than DP.

\subsection{Anomaly Detection}
Anomaly detection is crucial in safeguarding FL processes by identifying irregular updates and thwarting malicious influences. Collaborative training renders FL models susceptible to malevolent manipulations like model and data poisoning. These attacks have given rise to innovative FL techniques known as poisoning MIAs~\cite{NMS+19, TFS+22, CYS+22}, which exploit vulnerabilities to enhance the leakage of membership information. 
Notably, the poisoning MIAs focus on sample-level privacy leakage, diverging from conventional model poisoning attacks that seek to compromise overall model performance. Robust aggregation algorithms such as Multi-Krum~\cite{BPE+17}, Trimmed-Mean~\cite{YDC+18} and FLTrust~\cite{CXF+20} are inadequate in countering this active threat~\cite{MMZ+23}. 
Consequently, the anomaly detection approach are tailored to combat active MIAs in FL.

Ma et al.~\cite{MMZ+23} introduce an innovative client-side countermeasure called LoDEN to defend against active attacks in~\cite{NMS+19} by identifying and removing suspicious training samples. The active attack, executed through the gradient ascent algorithm, deliberately alters the model update of specific training examples, allowing the attacker to infer the membership status by observing gradient changes.
LoDEN employs a localized approach to counteract its impact on the FL model.
It identifies malicious updates by monitoring abrupt changes in the model outputs of training samples. 
Specifically, if the predicted label for a training sample shifts from correct to incorrect over several iterations, LoDEN identifies it as a malicious example. This example and its neighbors are removed from subsequent training, thus safeguarding the membership privacy of target models. 

\begin{table*}[htbp]
	\setlength\tabcolsep{2.0pt}
	\footnotesize
	\centering
	\caption{Summary of studies on membership inference defenses in federated learning.}
	\label{tab:defense2}
	\begin{threeparttable}
	\begin{tabular}{lllccll}
		\toprule
		\textbf{Type}  & \textbf{Year} & \textbf{\makecell{Reference}} & \textbf{Defender} &\textbf{\makecell{Corres. \\ attack}} & \textbf{\makecell{Defense \\ approach}}  & \textbf{\makecell{Comparison}}  \\
		\hline
		\midrule 
		\multirow{6}{*}{Partial sharing} & 2019 & Melis et al. \cite{MLS+19}   & Server & \cite{MLS+19} & Selective gradient &  -  \\
		\cline{2-7}
		~ & 2018 & Bernstein et al.~\cite{BJZ+18}   & \makecell{Client \\ Server} & \cite{JLN+23} & Gradient sign & \makecell{Differential privacy \\ Mix-up+MMD~\cite{LJL+20}} \\
		\cline{2-7}
		~ & 2019 & Chang et al.~\cite{CHS+19}   &  Server & \cite{NMS+19} & Prediction aggregation & - \\
		\cline{2-7}
		~ & 2022 & Stripelis et al.~\cite{SDG+22}   & Server & \cite{GUS+21} & Weight pruning & FedAvg  \\
		\midrule
		\multirow{2}{*}{Secure aggregation} & 2017 & Bonawitz et al.\cite{BKI+17}   & Client & \cite{GYB+22} & \makecell[l]{Secret sharing \& \\ Double-masking} & - \\
		\cline{2-7}
		~ & 2017 & Bai et al.\cite{BYF+21}   & Client & \cite{NMS+19} & HE & - \\
		
		\midrule
		\multirow{10}{*}{Noise perturbation} & 2018 & Rahman et al.~\cite{RMA+18}   & \makecell{Client \\ Server} & \cite{SRM+17} & LDP & Random guessing \\

		\cline{2-7}
		~ & 2021 & Hu et al.~\cite{HHS+21}   &  Client& \cite{HHS+21} &LDP  & -  \\

		\cline{2-7}
		~ & 2022  & Suri et al.~\cite{SAK+22}  & Client & \cite{SAK+22} & \makecell[l]{LDP \\ Subject DP } & Random guessing  \\
		
		\cline{2-7}	
		~ & 2022  & Liu et al.~\cite{LZL+22} & Server & \cite{SRM+17} & \makecell[l]{LDP \& \\ Trust domain}  &- \\
		
		\cline{2-7}
		~  & 2022 & Naseri et al.~\cite{NNJ+22}  &  \makecell{Client \\ Server}  & \cite{NMS+19} & \makecell[l]{LDP \\ CDP } & Norm bounding \\
		
		\cline{2-7}
		~ & \multirow{2}{*}{2022} & \multirow{2}{*}{Yang et al.~\cite{YXF+22}}  & \multirow{2}{*}{Server} & \multirow{2}{*}{\cite{NMS+19}} & \multirow{2}{*}{Random perturbation} & \multirow{2}{*}{\makecell[l]{FedAvg, PPDL~\cite{SRZ+15} \\ DBCL~\cite{FLN+20}, SPN~\cite{ZMW+21}}} \\
		& & & & & & \\
		\cline{2-7}
		~ & 2021  & Xie et al.~\cite{XYC+21} & Server & \cite{NMS+19} & Adversarial example  & Random guessing \\	
		\cline{2-7} 
		~ & 2021 & Lee et al.~\cite{LHK+21}   & Client & \cite{NMS+19} & Digestive network & DP  \\
		
		\midrule
		Anomaly detection & 2023 & Ma et al.~\cite{MMZ+23}   & Client & \cite{NMS+19} & Sample selection & Robust aggregation \\	
		\bottomrule	
	
	\end{tabular}
	\end{threeparttable}
\end{table*}

\begin{table*}[htb]
	\setlength\tabcolsep{6.0pt}
	\footnotesize
	\centering
	\caption{A comparison of defense approaches in centralized and federated learning.}
	\label{tab:comp-defense}
	\begin{threeparttable}
		\begin{tabular}{c|c|c|c|c}
			\toprule
			\bf{Setting} & \bf{Type} & \bf{Phase} & \bf{Protection}  & \bf{Unique}\tnote{1}\\
			\hline
			\hline
			\multirow{3}{*}{CL} & Output perturbation & Inference & Model output  & \CIRCLE \\
			\cline{2-5}
			~ & Regularization & Training & Target model  & \LEFTcircle  \\
			\cline{2-5}
			~ & Knowledge distillation & Training & Target model  & \CIRCLE  \\
			\cline{2-5}
			~ & Differential privacy  & Training & Target model & \Circle  \\
			\hline
			\hline
			\multirow{4}{*}{FL} & Partial sharing & Communication & Model update  & \CIRCLE  \\
			\cline{2-5}
			~ & Secure aggregation  & Communication and aggregation & Model update and itself  & \CIRCLE  \\
			\cline{2-5}
			~ & Noise perturbation  & Training and aggregation & Target model  & \LEFTcircle  \\
			\cline{2-5}
			~ & Anomaly detection & Aggregation & Target model  & \CIRCLE  \\
			
			\bottomrule
		\end{tabular}
		\begin{tablenotes}
			\item[1] \CIRCLE: unique defense \LEFTcircle: partially unique defense \Circle: common defense
		\end{tablenotes}
	\end{threeparttable}
\end{table*}

\subsection{Comparison to Defenses in Centralized Learning}
This section discusses four strategies to mitigate MIAs in FL, but their capabilities to protect against attacks vary.
Partial sharing hides certain updates to diminish the effectiveness of attacks, which can be applied to arbitrary threat models.
Regarding secure aggregation utilizing cryptographic techniques, SMC and HE generally provide defense against server-side attacks but become ineffective to the inference performed by a client with legal access to the global model.
One of the noise perturbation methods, DP offers a strict theoretical guarantee in safeguarding against attacks from all sides in an arbitrary strategy but with an inevitable utility loss, which motivates researchers to address this limitation by incorporating removable or well-crafted noises.
The anomaly detection method is tailored to counter active attacks but proves ineffective against passive attacks.
Table~\ref{tab:defense2} summarizes representative defense studies that have reported empirical performance in terms of release year,  defender role, defense approach, corresponding MIAs (i.e., corres.attack), and comparison methods.

Moreover, our research explores a comparative analysis of mitigation strategies in CL and FL, with the goal of deepening the understanding of countermeasures specifically within the FL context.
We examine defensive strategies from two angles: the phase when a defender implements countermeasure algorithms and the objects they aim to safeguard against information leaks. These comparisons are outlined in Table~\ref{tab:comp-defense}.
Notably, the process of model updates, a distinctive trait within the FL framework, stands out as an additional avenue for potential breaches in membership privacy. Consequently, defense strategies employed in FL, such as partial sharing, secure aggregation, and anomaly detection, significantly differ from those related to CL contexts.

\section{Future Direction}
\label{sec:future}

The field of MIA privacy is rapidly evolving, presenting numerous challenges and opportunities.
In the following, we discuss potential research directions on MIAs and defense strategies in the context of FL.

\subsection{Research Direction about Membership Inference Attacks}
\textbf{Holistic Evaluation Metrics.}
The existing evaluation metrics for MIAs provide an incomplete picture of attack performance~\cite{RSL+21}.
Although previous attacks perform well in the FL setting, their evaluation is focused only on member classes, such as the attack accuracy \cite{NMS+19, MLS+19, HHS+21}, precision, and recall \cite{ZJZ+20, ZYC+21}.
Such evaluations are often misleading owing to high false positive rate (FPR) or false alarm rate (FAR), which inspires researchers to consider the performance of negative samples by adding FAR or TPR at low FPR \cite{RSL+21, CNC+22}.
Furthermore, current metrics provide an average evaluation of record-level membership, which is unsuitable for source-level MIAs due to the non-IID phenomenon in FL.
Hence, it is necessary to develop a comprehensive and fair evaluation metric for various MIAs.

\noindent \textbf{Extension to Realistic Assumptions.}
Current strategies for MIAs in FL typically involve strict assumptions that may not hold in real-world scenarios, including the presence of IID training members~\cite{JLN+23}, a limited number of participants \cite{MLS+19, NMS+19, JLN+23}, and all federation round joining \cite{ZJZ+20, ZYC+21}.
These unrealistic assumptions lead to a misunderstanding of the effectiveness of previous attacks.
For example, heterogeneous data distributions are associated with a higher inference risk than IID data~\cite{SAK+22}.
Moreover, different epochs in which participants join FL influence the performance of MIAs for a local adversary~\cite{NMS+19}.
Future development should consider relaxing these assumptions and implementing MIAs in real-world and dynamic scenarios.

\noindent \textbf{Attacks toward Emerging Frameworks}.
The membership privacy risks associated with emerging FL frameworks remain underexplored.
Although several researchers have focused on centralized FL \cite{MBM+17, HAR+18}, in which a central server performs aggregations and broadcasts the aggregated results to participants, research on decentralized FL, such as peer-to-peer \cite{LQZ+20, ZLN+18} and blockchain \cite{WRL+21, ZYZ+20}, remain limited.
In the case of decentralized frameworks, a curious attacker can observe the latest model updated by a known client, thereby facilitating the launch of source-level MIAs.
Furthermore, when it comes to widely-used VFL framework~\cite{GBD+20}, previous MIAs prove to be generally ineffective due to the fact that each participant possesses only a portion of the feature space and has access to an incomplete target model.

\noindent \textbf{Emsembled Attack Strategies.}
Considering that FL is susceptible to various security attacks, such as poisoning attacks \cite{BPE+17, BAN+19, FMC+20, CDC+19, WLY+24} and backdoor attacks \cite{BEV+20, XCH+20, CCG+20, LHY+23}, there is an opportunity for future work to incorporate these attacks to enhance the effectiveness of existing MIA techniques or develop potent attack approaches.
In particular, data poisoning techniques have emerged as a significant concern, as they can significantly increase the privacy risks associated with benign training samples and amplify the membership exposure of the targeted class effectively \cite{CYS+22, ZYB+23, TFS+22}.

\noindent \textbf{Reasons for Information Leakage.}
The comprehensive analysis of membership information leakage in the context of FL remains inadequate.
In the CL scenario, there is considerable evidence that overfitting facilitates the success of MIAs in the black setting \cite{SRM+17, YSG+18, SAM+19, SLM+21, SLS+19-1}.
In contrast, regarding white-box FL models, internal exposure, non-IID training samples, and cooperative learning offer adversaries more opportunities~\cite{NMS+19}. They present challenges in exploring the underlying reasons behind these factors theoretically and empirically.
Such investigations can uncover vulnerabilities within the FL framework and advance the development of effective attack approaches.

\subsection{Research Direction about Membership Inference Defenses}

\textbf{Unified Evaluation Benchmark.}
There is a lack of a standardized evaluation benchmark for assessing the effectiveness of defense mechanisms.
Currently, countermeasures are typically evaluated based on distinct MIAs \cite{SRZ+15, HHS+21, SAK+22} or discussed generally without experimental evidence \cite{SS+20, FHM+21}.
Although several researchers \cite{CHS+19, YXF+22, XYC+21, NNJ+22} have evaluated the effectiveness of defense strategies for the same MIAs~\cite{NMS+19}, their results cannot be compared owing to the use of different datasets and networks.
Consequently, it is challenging to identify suitable mitigation mechanisms for a defender.
A unified evaluation benchmark can address this problem and provide valuable guidance in making practical choices.

\noindent \textbf{Defense Algorithm Assessment.}
Previous research has often overstated the effectiveness of defense mechanisms in mitigating information leakage.
Non-IID data and homogeneous model architecture have a significant impact on privacy leakage risk. Unfortunately, they are often overlooked when designing new defense mechanisms \cite{SAK+22, GUS+21, YWY+23}.
Secure aggregation is often considered an effective technique for an honest-but-curious server. However, recent research~\cite{GJH+21} argued that a server could infer categorical information about a specific participant.
Indeed, it indicates that when the server acts as an attacker, relying on secure aggregation in FL is insufficient to protect membership information adequately.
As a result, there is a growing need for researchers to evaluate these mechanisms under realistic assumptions and attack scenarios.

\noindent \textbf{Privacy-Utility-Efficiency Defense Mechanisms.}
There is a scarcity of effective countermeasures that guarantee privacy preservation with small utility losses and low computational costs.
DP mechanisms provide strong privacy guarantees at the cost of large utility loss \cite{RMA+18, JBE+19, NNJ+22} and are thus unsuitable for mission-critical applications.
Partial sharing mechanisms only slightly decrease the effectiveness of attack performance \cite{MLS+19, JLN+23}, and secure aggregation suffers from high computational costs.
Existing defense mechanisms fall short of providing comprehensive protection for membership information. This issue could be addressed by integrating multiple privacy preservation methods, leading to enhanced privacy safeguards \cite{LYK+22, SRZ+15}.

\noindent \textbf{Robust Defense Mechanisms.}
Exploration of robust protection strategies shows promise in preventing membership leakage while avoiding introducing additional risks.
An instance of this is when FL protected by DP inadvertently creates an opportunity for model poisoning attacks, enabling attackers to evade anomaly detection~\cite{YMC+23}.
Additionally, given the strong negative correlation between MIAs and model extraction attacks \cite{TFZ+16, LYW+22}, the mitigation of MIAs may improve the effectiveness of model extraction attacks.
Such occurrences should be avoided as the objective is to build a robust and safe FL model in most scenarios.
Hence, researchers should consider the latent correlation among attacks when developing defense approaches against MIAs.

\section{Conclusion}
\label{sec:concl}
MIAs represent a critical and rapidly evolving research area within FL.
This survey comprehensively summarizes the landscape of MIAs and corresponding defenses within the FL framework, providing structured taxonomies to categorize existing literature. Based on the utilization of attack knowledge in these studies, we categorize MIA research into two primary approaches: update-based and trend-based.
In addition to attacks in FL, we highlight prevalent defense mechanisms currently deployed to mitigate the vulnerabilities MIAs pose. These defenses encompass a range of strategies, from partial sharing techniques to noise perturbation.
Furthermore, this survey identifies promising avenues for future research in MIA and defense strategies. Key opportunities include exploring novel attack vectors in diverse FL settings, refining existing defense mechanisms, and integrating robust privacy-preserving techniques into FL frameworks. 
These future studies can help enhance security and privacy guarantees in FL.

\begin{acks}
	This work was supported by the National Natural Science Foundation of China (Grant No: 62072390, 92270123, 62102334), and the Research Grants Council, Hong Kong SAR, China (Grant No:
	15222118, 15218919, 15203120, 15226221, 15225921, 15209922, and C2004-21GF).
\end{acks}

\bibliographystyle{ACM-Reference-Format}
\bibliography{sample-base}

\end{document}